\documentclass[useAMS,usegraphicx,usenatbib,float]{mn2e}
\usepackage{color}
\usepackage{times}
\usepackage[dvips]{graphicx}
\usepackage{amsmath,amssymb,graphicx}
\usepackage{mathrsfs}
\newcommand{\sm}[1]{{\textcolor{black}{#1}}}
\newcommand{\fig}[1]{{\textcolor{black}{#1}}}

\newcommand{\simgt}{\lower.5ex\hbox{$\; \buildrel > \over \sim \;$}}
\newcommand{\simlt}{\lower.5ex\hbox{$\; \buildrel < \over \sim \;$}}
\def\hkpc{h^{-1}{\rm kpc}}
\def\hMpc{h^{-1}{\rm Mpc}}
\def\hMsun{h^{-1}M_\odot}
\title[Modeling color-dependent galaxy clustering]
{Modeling color-dependent galaxy clustering in cosmological simulations}

\author[S. Masaki, Y.-T. Lin \& N. Yoshida]{{Shogo Masaki$^{1,2}$\thanks{E-mail: shogo.masaki@nagoya-u.jp}\thanks{JSPS Fellow}, Yen-Ting Lin$^{3,4}$ and Naoki Yoshida$^{4,5}$}\\
$^1$Department of Physics, Nagoya University, Nagoya, Japan\\
$^2$NTT Secure Platform Laboratories, NTT Corporation, Musashino, Japan\\
$^3$Institute of Astronomy and Astrophysics, Academia Sinica, Taipei, Taiwan\\
$^4$Kavli Institute for the Physics and Mathematics of the Universe (WPI), TODIAS, The University of Tokyo, Kashiwa, Japan\\
$^5$Department of Physics, The University of Tokyo, Tokyo, Japan}
\begin{document}

\date{Accepted . Received ; in original form }

\pagerange{\pageref{firstpage}--\pageref{lastpage}} \pubyear{}

\maketitle

\label{firstpage}
 
\begin{abstract}
 We extend the subhalo abundance matching method to assign galaxy color to subhalos.
We separate a luminosity-binned subhalo sample into two groups by a secondary subhalo 
property which is presumed to be correlated with galaxy color. 
The two subsamples then represent red and blue galaxy populations.
We explore two models for the secondary property, \sm{namely} subhalo assembly time and 
local dark matter density around each subhalo.
The model predictions for the galaxy two-point correlation functions 
are compared with the recent results from the Sloan Digital Sky Survey.
We show that the observed color dependence of galaxy clustering can be reproduced well
by our method applied to cosmological $N$-body simulations without baryonic \sm{processes}.
We then compare the model predictions for the color-dependent galaxy-mass cross 
correlation functions with the results from gravitational lensing observations.
The comparison allows us to distinguish the models, and also 
to discuss what subhalo property should be used to assign color to subhalos accurately.
We show that the extended abundance matching method using the local dark matter density 
as a color proxy provides an accurate description of the galaxy populations in the local universe.
\sm{We also study impacts of scatter in the local dark matter density-color relations.
Introducing scatter improves agreements of our model predictions with the observed red and blue galaxy clustering and is needed to explain observed correlation functions in finer color bins.
Finally, we study red galaxy fraction profiles in galaxy group- and cluster-sized halos and find the red fraction profiles have a relatively strong dependence on our model parameters.
We argue that the red fraction profiles can be an important observational clue, in addition to galaxy clustering and galaxy-galaxy lensing, to explore the galaxy-(sub)halo connections.}
\end{abstract}


\section{Introduction}
Describing galaxy formation and evolution within the context of
the standard cosmology is one of the most important goals in astronomy and cosmology.
Large data \sm{sets} from recent galaxy redshift surveys, such as the Sloan Digital 
Sky Survey \citep[SDSS;][]{York00}, 
have enabled us to measure precisely various statistics \sm{even} for subsamples of 
the galaxies classified by, for example, 
luminosity, stellar mass, morphology and so on \citep[e.g.,][]{Li06,Zehavi11}.
The upcoming deep and wide-area galaxy surveys, 
e.g., Subaru Hyper Suprime-Cam Survey
\footnote{http://subarutelescope.org/Projects/HSC/ also http://sumire.ipmu.jp/en/} 
and Dark Energy Survey\footnote{http://www.darkenergysurvey.org/}, will provide
even larger data of distant galaxies.
Clearly, accurate theoretical models are needed to interpret 
observational results such as those on the relationship between galaxies 
and dark matter (DM) halos.

So far, several useful phenomenological methods to link halos with galaxies have 
been developed. One is the so-called halo occupation distribution (HOD) 
modeling \cite[see e.g.,][]{Seljak00,CooraySheth02,Tinker05,Blake08,Zehavi11,Leauthaud12}.
The method parametrizes HODs, i.e., occupation number of galaxies per halo, of central 
and satellite galaxies separately, as a function of halo mass.
The functional form of HOD is inspired by hydrodynamical cosmological simulations 
or by semi-analytic models \citep{Zheng05}.
By using the halo mass function as well as the bias factor and the DM density profile 
obtained from cosmological $N$-body simulations, 
one can \sm{easily} calculate the galaxy two-point correlation functions \sm{analytically}.

There are free parameters in the HOD formalism which are constrained by matching to the observed 
clustering \sm{properties} and to the number density of galaxies \citep{Blake08,Zehavi11,Coupon12}.
More sophisticated treatments of HOD include the conditional luminosity functions 
\citep{vdB03,Cooray06}. An attractive feature of HOD is that it can be extended to model
and calculate \sm{not only galaxy clustering but also} other statistical quantities \citep[e.g.,][]{DeBernardisCooray12,Leauthaud12,MasakiYoshida12,vdB12}.
However, it is known that one often encounters difficulty in fitting HOD model predictions 
to the observed clustering and the number density simultaneously 
\citep{Quadri08,Matsuoka11,Wake11}.

The second method is called the subhalo abundance matching \citep[SHAM;][]{Kravtsov04,Tasitsiomi04,Conroy06,Guo10,Moster10,WetzelWhite10,Neistein11,Trujilo-Gomez11,Hearin12,Masaki13,Reddick12,Rodriguez-Puebla12,HearinWatson13}.
SHAM assumes a tight relation between galaxy luminosity (or stellar mass) 
and subhalo circular velocity (or mass).
The threshold circular velocity is set such that the number density of 
subhalos and that of luminosity-selected galaxies are matched as
\begin{eqnarray}
  n_{\rm subhalo}(>V_{\rm cir})=n_{\rm galaxy}(>L).
  \label{sham}
\end{eqnarray}
The predicted clustering generally agrees with observations,
reproducing the observed luminosity dependence of galaxy clustering.
It should be emphasized that SHAM does not need fitting parameters while HOD modeling does.
An additional advantage of SHAM is that it can utilize $N$-body simulation outputs directly.
Therefore, SHAM can incorporate the non-linear evolution of galaxy clustering accurately 
while having a simple and direct galaxy-subhalo connection.

Both the HOD model and SHAM are used to calculate the galaxy two-point correlation 
or the galaxy-mass cross correlation as a function of a galaxy property, 
e.g., luminosity and stellar mass. 
It would be ideal if multiple observable properties are assigned simultaneously 
to (sub)halos. 
Galaxy color is one of the most fundamental properties. 
It is thought to indicate the galaxy's age and the level of star formation activity.
Also it is well-known that there are two apparent sequences in the color-magnitude 
diagram, which represent red and blue galaxies. The two populations differ in 
many aspects including their spatial clustering, reflecting probably their
formation histories.
So far, detailed treatments for color-dependent clusterings or lensing measurements 
have been developed, but \sm{mostly} within the HOD model 
\citep{vdB03,Cooray06,Mandelbaum06,Tinker08,RossBrunner09,Simon09,SkibbaSheth09,Zehavi11}.

In the present paper, we develop a method to assign galaxy color to subhalos by extending SHAM.
We first apply SHAM to a subhalo catalog to obtain luminosity-selected subhalo samples. 
We then divide \sm{a magnitude binned subhalo sample} into two groups by ordering 
subhalos using a secondary quantity.
The two groups are then meant to represent red and blue galaxy samples.
Similarly to the original SHAM, 
the abundance ratio of the divided two groups is matched to the observed red/blue ratio.
Clearly, we need to choose an appropriate subhalo property which is presumably 
correlated with galaxy color. 
We propose two models for the secondary property.
One is motivated by the so-called assembly bias whereas 
the other incorporates environment effects.

Assembly bias is a property-dependent bias for halos in a fixed mass bin. 
A variety of properties have been considered so far, 
such as assembly time, spin, concentration and so on \citep[see e.g.,][]{Gao05,Wechsler06,Croton07,GaoWhite07,Reed07,LacernaPadilla11,LacernaPadilla12}.
Among these, we choose subahlo age as a proxy of galaxy color.
\sm{After the submission of this paper, \cite{HearinWatson13} presented an approach that is similar in spirit to our method in assigning colors to subhalos.
They showed that the color dependences of galaxy clustering can be explained by using the redshift at which the galaxy is starved of cold gas as a color proxy.}

It is thought that there are some environmental effects through which
a galaxy's morphology and color evolve.
\cite{Gerke12} recently developed a model to assign color to subhalos \cite[see also][]{Tasitsiomi04}.
In their model, color is assigned to subhalos by using local galaxy density 
along with the empirical color-galaxy density relation 
\citep[e.g.,][]{Hogg04,Balogh04,Bamford09}.
Our second model is similar to the method in \cite{Gerke12},
although we do not use such 
an empirical relation. 

We consider lensing measurements in addition to galaxy clusterings
in order to distinguish the two models. Lensing observations will provide an
independent information which possibly allows us to derive an accurate galaxy-halo 
connection. In particular, we utilize the observed 
early/late-type dependent lensing profile \citep{Mandelbaum06}
assuming early/late-type galaxies approximately correspond to
red/blue galaxies.
We show, for the first time, that an extended 
SHAM can reproduce the observed lensing profiles as well as
the color-dependent galaxy clusterings in the local universe \sm{by using the local DM density as a color proxy.}

\sm{Finally, we introduce scatter in the local DM density-color relations by perturbing the originally measured local density.
We discuss impacts of scatter on galaxy clustering measurements for not only red and blue galaxy samples but also subsamples of finer color binned.
We also study the relative proportion of red vs blue galaxies as a function of distance from the halo center for group- and cluster-scale halos.}

The rest of the present paper is organized as follows.
In Section 2, we describe our cosmological $N$-body simulations and the basics of SHAM.
In Section 3, we explain our method of galaxy color assignment.
Then in Section 4, we compare the model predictions for color dependent 
galaxy clustering and galaxy-mass cross correlations with the observational results and present the results for the impacts of scatter in the local DM density-color relations.
We summarize and discuss our results in Section 5.

\section{Setup}
\subsection{Cosmological $N$-body simulations}
We use the massively parallelized code {\it Gadget-2} \citep{Springel01a,Springel05} 
in its Tree-PM mode to run two cosmological $N$-body simulations.
For each run, the assumed $\Lambda$-cold-DM cosmology is consistent with the {\it Wilkinson Microwave Anisotropy Probe (WMAP)} 7-year results, with $\Omega_m=0.272, \Omega_b=0.0441, \Omega_\Lambda=0.728, H_0=100h=70.2~{\rm km~s^{-1}~Mpc^{-1}}, \sigma_8=0.807$ and $n_s=0.961$ in the standard notation \citep{Komatsu11}.
The two runs, one with $200~\hMpc$ on a side and the other with $300~\hMpc$ box size, both employ $1024^3$ DM particles, and are referred to as L200 and L300, respectively.
The resulting mass of a particle is $5.6\times10^8~\hMsun$ ($1.9\times10^9\hMsun$) for the L200 (L300) run.
We use the code {\it CAMB} \citep{CAMB} to obtain the matter power spectrum for the initial condition.
We set the initial redshift to be 49 (65) for the L200 (L300) run.
\sm{The initial conditions for both runs are generated using the second-order Lagrangian perturbation theory \citep{Crocce06,Nishimichi09}.}
The gravitational softening length is set to be $5~\hkpc$ and $8.8~\hkpc$ for the L200 and the L300 run, respectively.
The L300 run has lower mass and spatial resolution than the L200 run but provides higher statistical precision of correlation functions for massive objects and at the large scale.

\subsection{Identification of halos, tracking assembly history and construction of subhalo catalog}
We identify the distinct halos, i.e., halos that do not lie 
in more massive halos, using the friends-of-friends (FoF) algorithm \citep[e.g.,][]{Davis85} with the linking parameter 
of 0.2 in units of the mean interparticle separation.
For the identification of subhalos, i.e., dense self-gravitating clumps 
that reside in a distinct halo, we utilize the {\it SubFind} algorithm 
developed by \cite{Springel01b}.
The algorithm decomposes a distinct halo into a central subhalo, i.e., 
the so-called smooth component which contains the majority of mass, 
and satellite subhalos.
We store the subhalos with more than 20 particles.

To implement SHAM, we construct the subhalo catalog with $V_{\rm max}^{\rm acc}$ for the 
satellite subhalos, where $V_{\rm max}^{\rm acc}$ is the maximum value of particle circular 
velocity $V_{\rm cir}(r)=\sqrt{GM(<r)/r}$ at the last epoch when the satellite subhalo was a central one \citep{Conroy06}.
We follow the recipe described in detail by \cite{Allgood05} to construct the mass assembly history 
of the most massive progenitor of subhalos.
We use 50 simulation snapshots taken evenly in $\ln(1+z)$ from $z=10$ to $z=0$ for the two simulations.
For the central subhalos, we tabulate the maximum circular velocity 
at the redshift when they are {\it observed}, $V_{\rm max}^{\rm now}$.
Then the term $V_{\rm cir}$ in Equation (\ref{sham}) is replaced with $V_{\rm max}^{\rm now}$ for the central subhalos 
and $V_{\rm max}^{\rm acc}$ for the satellite subhalos.
Note that the circular velocity defined in this way is a direct proxy of gravitational potential, and is less sensitive to subhalo identification algorithms than is the subhalo mass.

\subsection{Application of SHAM}
Following \cite{Zehavi11}, we consider \sm{the SDSS galaxies at $z\simeq0$} in three magnitude bins, 
$-22<M_r-5\log_{10}h<-21,~-21<M_r-5\log_{10}h<-20$ and $-20<M_r-5\log_{10}h<-19$, and hereafter denote the absolute magnitude just as $M_r$ without the $-5\log_{10}h$ term.
We apply SHAM to the subhalo catalogs constructed in the L200 and L300 runs to create the magnitude-binned subhalo samples.
We take the bracketing threshold samples for a bin and then use the difference of them as a binned sample.
The comoving number density for each threshold sample can be found in \cite{Zehavi11}.

\begin{figure}
\includegraphics[width=8.5cm]{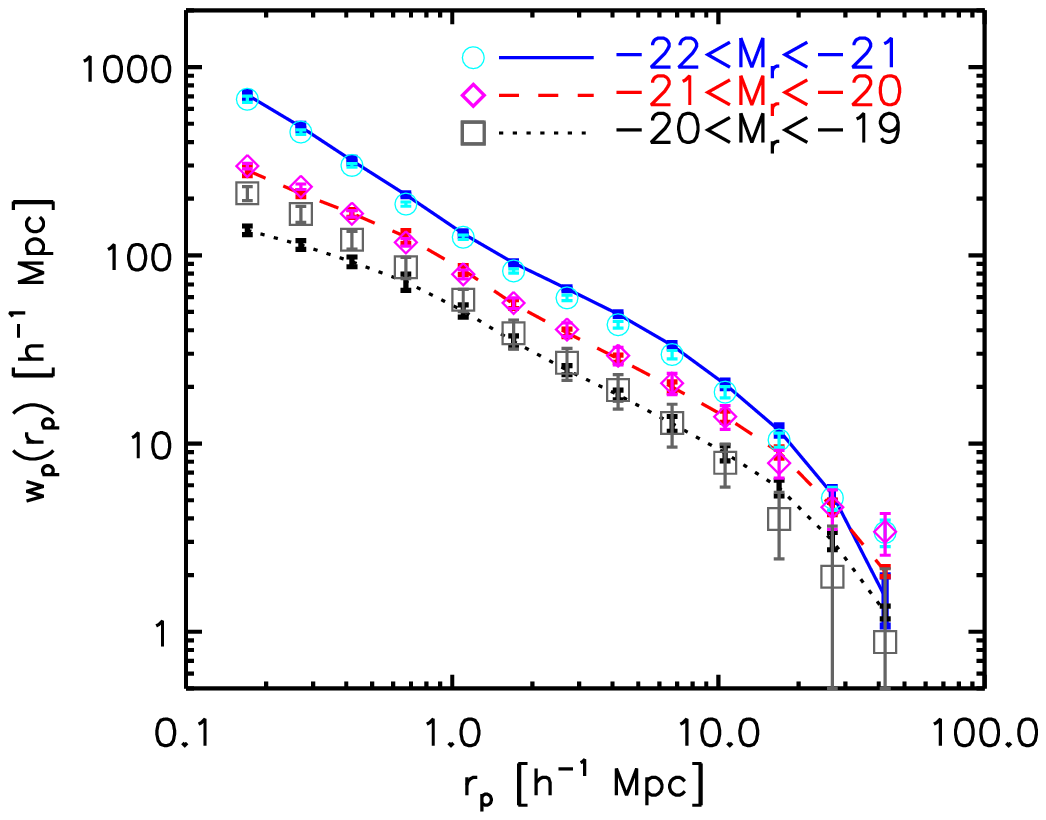}
\includegraphics[width=8.5cm]{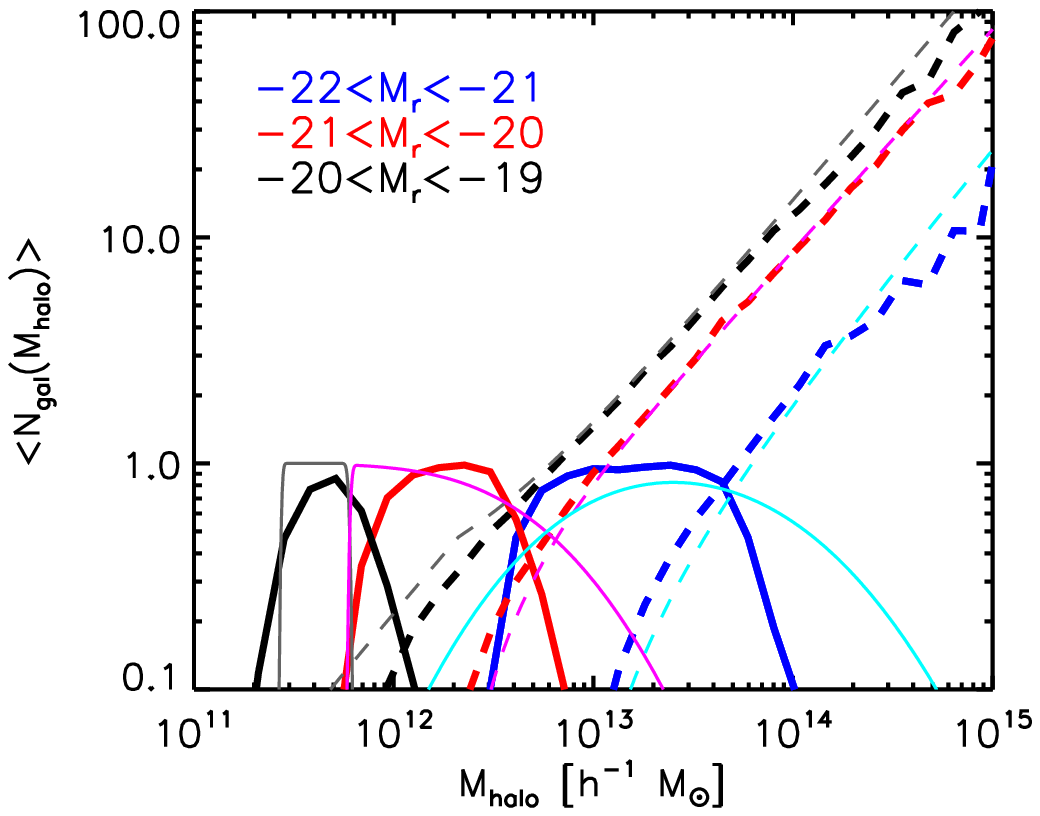}
\caption{{\em Top}: The projected correlation functions for the magnitude-binned samples. The SHAM predictions are shown by solid\fig{, dashed and dotted} lines with error bars calculated by jack-knife resampling of eight subvolumes. The SDSS results \citep{Zehavi11} are shown by open circles\fig{, diamonds and boxes} with error bars. The amplitudes of the brightest and the faintest sample are shifted upward and downward by $0.1~{\rm dex}$ for clarity. 
{\em Bottom}: HODs for each binned sample obtained by SHAM (the thick lines). For comparison, the best-fit models of HOD modeling \citep{Zehavi11} are shown by pale color thin lines. The solid and the dashed lines are HODs of central and satellite galaxies, respectively. Note that we quote the FoF mass as the distinct halo mass while the HOD modeling uses the virialized mass. Hence our HODs should be shifted to the left hand side slightly to be compared with the HOD fitting results.} 
\label{wp_hod_lum}
\end{figure}

We measure the projected correlation function $w_p$ as a function of the projected distance $r_p$ as
\begin{equation}
  w_p(r_p)=2\int_0^{\pi_{\rm max}} d\pi~\xi(r=\sqrt{\pi^2+r_p^2})
\end{equation}
\sm{where $r$ is the three dimensional distance, $\xi$ is the spatial two-point correlation function and $\pi$ is the distance along the line-of-sight.}
We take the same value of $\pi_{\rm max}$ used by \cite{Zehavi11} for each bin.
For the fainter and the intermediate samples we use the L200 subhalo catalog at $z=0$.
Only for the brighter sample, we use the catalog from the L300 run at $z=0.1$.
The chosen redshift is very close to the mean redshift of each magnitude binned galaxy sample.
The results are compared with \cite{Zehavi11} in Figure \ref{wp_hod_lum}.
The top panel shows fairly nice agreements between SHAM and the observation, as expected from previous works \citep{Conroy06,Trujilo-Gomez11}.
We found that SHAM overpredicts the amplitude for the brightest sample by about 10\% at all scales.
This may be because we do not include scatter in the $V_{\rm max}-M_r$ relation.
In reality, scatter between the two quantities is expected.
It has been pointed out that introducing scatter reduces the clustering amplitude, in particular for the brighter sample effectively, and makes the agreement better \citep{WetzelWhite10,Trujilo-Gomez11,Reddick12}.

We show the HODs obtained for each binned sample in the bottom panel and compare them with the best-fit HOD modeling results estimated by \cite{Zehavi11}.
We consider the galaxies assigned to the central and satellite subhalos as the central and satellite galaxies, respectively.
The figure shows that the HODs of central galaxies from SHAM are narrower than the HOD modeling for the bright and intermediate samples.
It should be noted that both HOD modeling and SHAM give very similar predictions for galaxy clustering.

\section{Methods for color assignment}
Here we describe our method to assign galaxy color to subhalos.
SHAM uses the abundance as a function of subhalo and galaxy properties to match their number densities.
SHAM works well if the properties, e.g., galaxy luminosity and the subhalo maximum circular velocity, highly correlate with each other \citep{Conroy06,Neistein11,Reddick12}.
The spirit of our method is based on this insight.
We separate each luminosity-binned subhalo catalog into two groups by a secondary subhalo property.
The secondary property is expected to be correlated well with galaxy color.
Then the two groups correspond to the red and blue galaxies within the luminosity bin.
In a similar fashion to SHAM, the number density ratio of the two subhalo groups is matched with the observed red/blue galaxy abundance ratio.
We study two models for the secondary subhalo property: one is motivated by assembly bias \cite[e.g.,][]{Gao05,Wechsler06,Croton07,GaoWhite07,Reed07,LacernaPadilla11} while the other incorporates environment in and around subhalos.
We discuss our two models below.

\subsection{Models for a proxy of galaxy color}
\subsubsection{Subhalo age}
It is naive but natural to consider that the galaxy color is a proxy of the galaxy age and to expect that subhalo age is highly correlated with the age of the galaxy\footnote{For simplicity, we do not consider bias in age due to starburst triggered by mergers.}.
We use the redshift evolution of the maximum circular velocity to define the formation epoch $z_{\rm form}$ of subhalos via
\begin{equation}
  V_{\rm max}(z=z_{\rm form})=f\times V_{\rm max},~~{\rm with}~0<f<1
  \label{zform}
\end{equation}
where $V_{\rm max}$ on the right hand side is $V_{\rm max}^{\rm acc}$ for the satellite subhalos and $V_{\rm max}^{\rm now}$ for the central subhalos at $z=0$ or $0.1$, depending on the simulation box \sm{or the magnitude sample.}
Note that $z_{\rm form}$ of the satellite subhalos is defined at an earlier epoch than the accretion epoch.
In practice, we identify the two snapshots between which $f\times V_{\rm max}$ is located and then we interpolate linearly between them to get $z_{\rm form}$.
For low mass subhalos or small values of $f$, we cannot follow down to $f\times V_{\rm max}$ due to the lack of mass resolution \sm{at high-$z$.}
In such cases, we define the redshift at which the subhalo is identified for the first time as $z_{\rm form}$.\footnote{As we will see later, high values of $f\simeq0.9$ are preferred to fit the observed clustering. For such $f$, the formation epochs of almost all subhalos are identified later than the first identified epoch even for the faint sample.}
We refer to this model as the {\it age model}.
In this model, the subhalos with higher $z_{\rm form}$ correspond to redder galaxies.
It has been shown extensively in the literature that the older distinct halos are more clustered than the younger ones \citep{Gao05,GaoWhite07,Reed07,LacernaPadilla11}.

\begin{figure}
  \includegraphics[width=8.5cm]{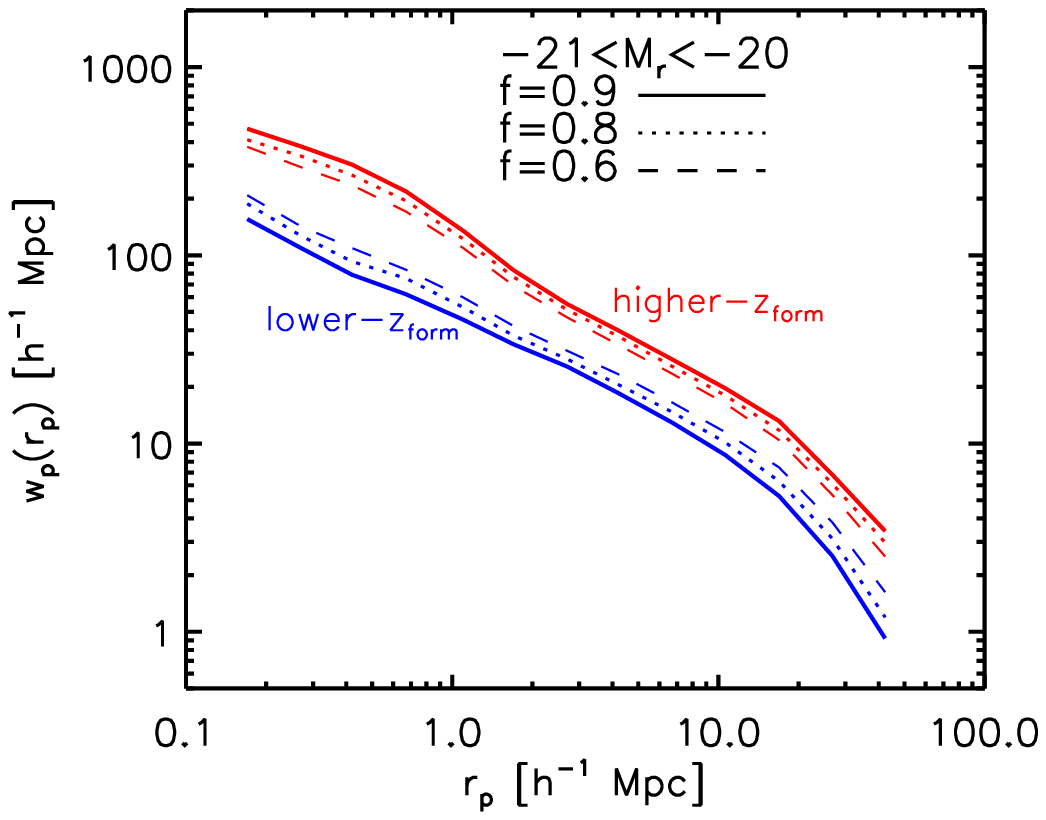}
  \includegraphics[width=8.5cm]{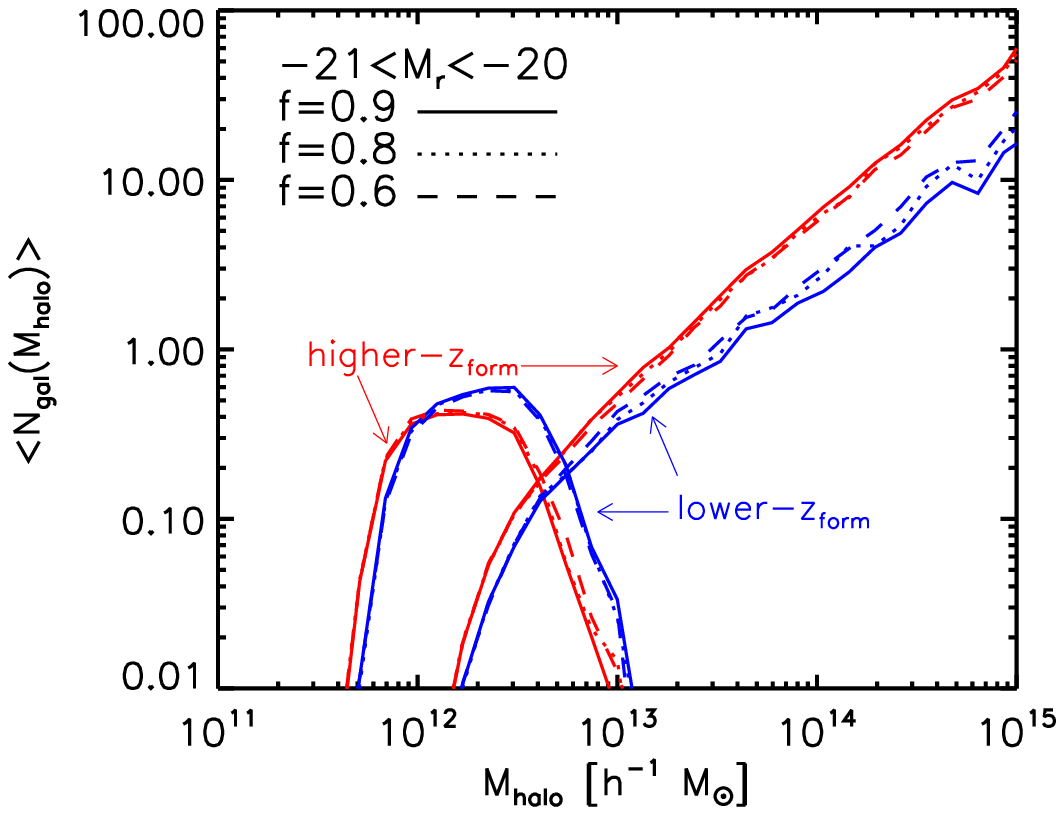}
  \caption{Impact of the adopted $f$ value (see Equation~[\ref{zform}]) on the clustering and HOD in our age model. The red and the blue lines show $w_p$ or HOD for the subhalo subsamples with higher- and lower-$z_{\rm form}$. The adopted values of $f$ are $0.9, 0.8$ and $0.6$. {\em Top}: Projected correlation functions from the age model. {\em Bottom}: HODs of central and satellite galaxies as a function of host halo mass.}
  \label{wp_hod_age}
\end{figure}

The choice of value of $f$ is not unique.
We  find that the predicted correlation function depends on the value of $f$.
We show the impact of $f$ in the age model in Figure \ref{wp_hod_age}, \sm{with $f=0.9,~0.8$ and $0.6$ for the $-21<M_r<-20$ sample.}
For the fainter and brighter samples, very similar trends are found.
The red and the blue lines represent $w_p$ or HOD for the subhalos with higher- and lower-$z_{\rm form}$, respectively.
As we expect, a ``split'' like the red/blue galaxy separation is seen in the correlation function (top panel).
Furthermore, a larger value of $f$ gives larger split.
This can be understood as a larger $f$ value captures subhalos which formed at earlier epoch.
Hence the predicted clustering for red galaxies is more enhanced \sm{for higher values of $f$}.
\footnote{\sm{Suppose two subhalos which have the same formation epoch if $f=0.6$, i.e., they had the circular velocity of $0.6\times V_{\rm max}$ at a redshift. 
After the redshift, one of them can be assembled at earlier epoch and evolve slower than another one.
The subhalo assembled earlier would be thought to be formed earlier and a more biased object.
The higher value of $f$ captures such situation.}}
In the bottom panel, HODs are shown.
The peaks at $M_{\rm halo}\simeq2\times10^{12}\hMsun$ are HODs of central galaxies and the power law-like curves at high halo mass regime are HODs of satellite galaxies.
It can be seen that the satellite HOD is more sensitive to $f$ than the central HOD.

\subsubsection{Local DM density around subhalos}
As our second model, we adopt the local DM density around subhalos 
as the secondary subhalo property.
\cite{Mandelbaum06} showed that early type galaxies have higher mass density profile than late type ones at $z=0\sim0.1$ in several magnitude bins via galaxy-galaxy lensing techniques.
In particular, this morphology dependence of mass density profile is more apparent at $r_p\ga100~\hkpc$.
This suggests that redder galaxies tend to be hosted by subhalos with higher density envelope than bluer ones.

We take a sphere with radius of $R_{\rm DM}$ centered on the center of a subhalo to measure the local DM density.
In this model, the subhalos with higher local density corresponds to the redder galaxy populations.
We take the sphere radius of $\sim\mathcal{O}(100~\hkpc)$ to reflect our motivation from the lensing study.
It should be noted that this property is less affected by simulation resolution than the age model.
We refer to this model as the {\it local density model}.

\begin{figure}
  \includegraphics[width=8.5cm]{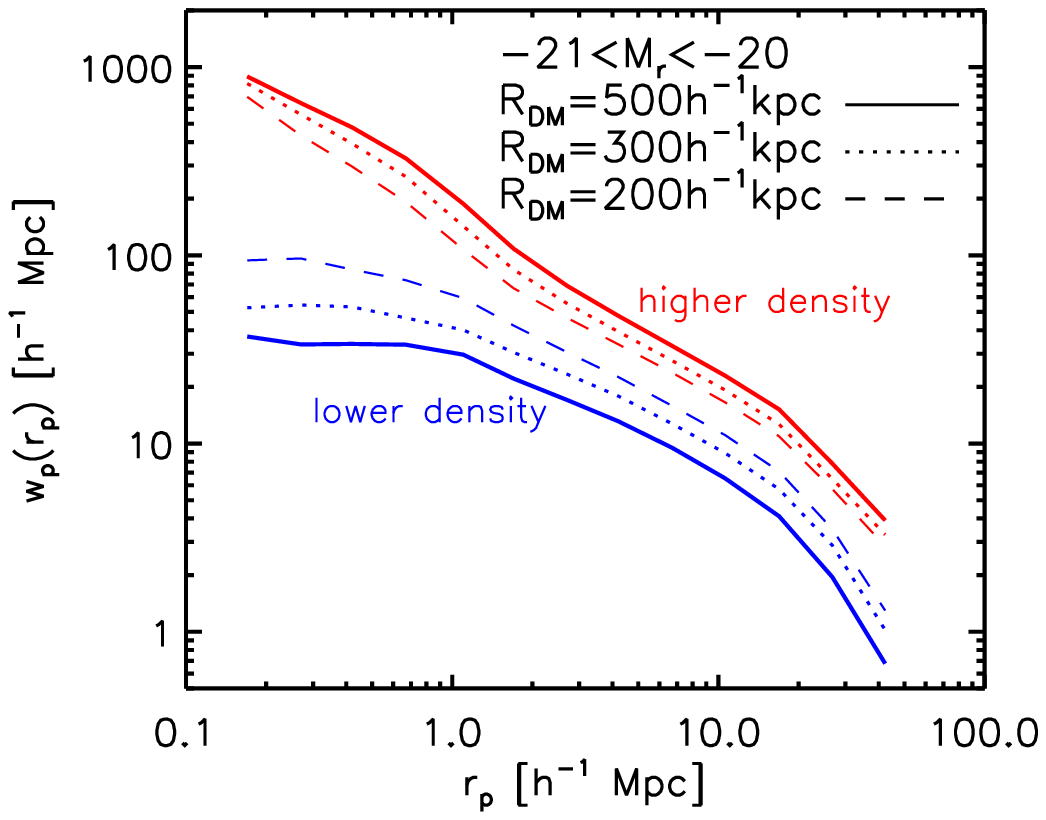}
  \includegraphics[width=8.5cm]{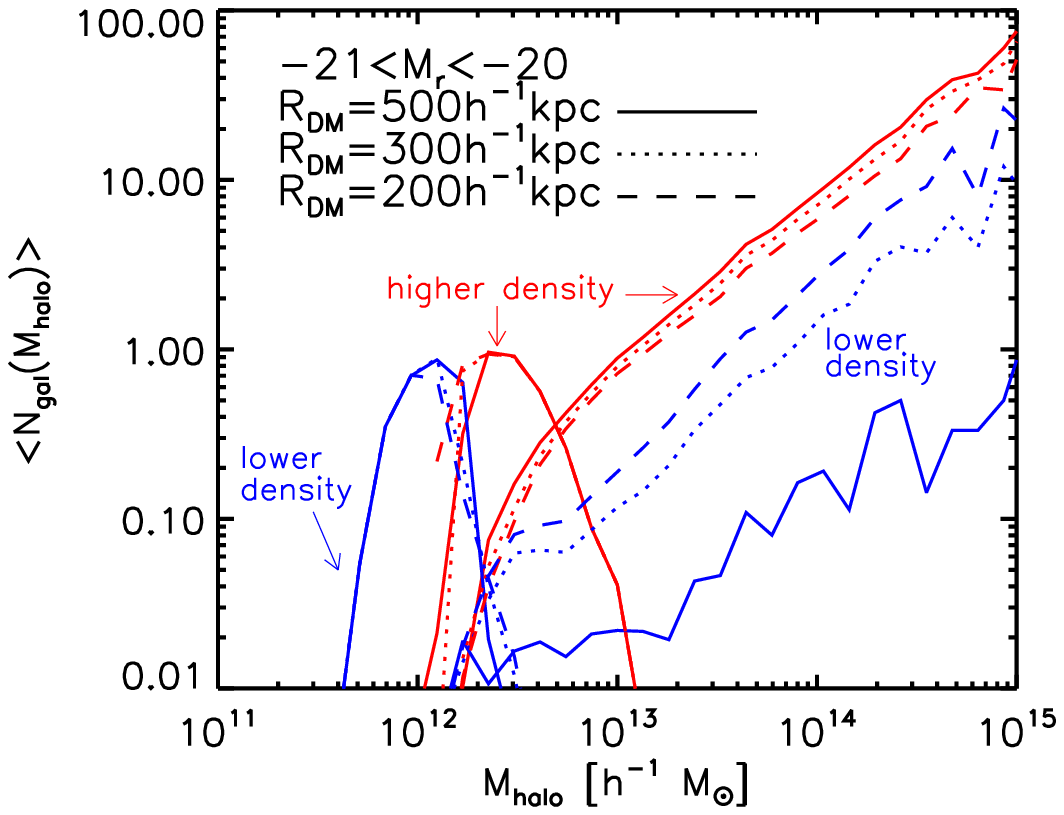}
  \caption{Same as Figure \ref{wp_hod_age} but for the local density model. The red and the blue lines show $w_p$ or HOD for the subhalo subsamples with higher and lower local density. The adopted sizes of $R_{\rm DM}$ are $500, 300 {\rm~and~}200~ \hkpc$.}
  \label{wp_hod_loc}
\end{figure}

Similar to the age model, we find that the predicted correlation function depends on the size of sphere $R_{\rm DM}$.
Figure \ref{wp_hod_loc} illustrates the impact of $R_{\rm DM}$ in the local density model.
The adopted sphere sizes in the figure are $R_{\rm DM}=500,~300$ and $200~\hkpc$.
Again, the predictions for the $-21<M_r<-20$ sample are shown.
The red and the blue lines are $w_p$ or HOD for the subhalos with higher and lower local DM density, respectively.
The top panel shows that clustering amplitudes for the subhalos with higher local density are higher than those for the subhalos with lower density as we expect.
The panel also shows that the color split is larger for larger-$R_{\rm DM}$.
These are because that counting simulation particles in a sphere with radius of $\ga100~\hkpc$ leads subhalos in massive halos (i.e., cluster- or group-sized halos) to be ones with higher local density.
This trend is more effective for larger-$R_{\rm DM}$ as shown in the HOD (see the bottom panel).
The bottom panel shows that the red central galaxies are assigned to higher mass distinct halos than the blue central galaxies.
This is because that our local density measure strongly traces subhalo mass or host distinct halo mass itself for central subhalos.

\section{Results}
\subsection{Color dependence of projected correlation functions}
\label{sec:wp}
We compare our model predictions for the color dependence of clustering with the observational results by \cite{Zehavi11}.
They separated the red and blue galaxies in the $(g-r)-M_r$ color-magnitude space with the division of 
\begin{equation}
  (g-r)=0.21-0.03M_r.
  \label{divider}
\end{equation}
For each magnitude bin, we search the model with the lowest value of $\chi^2=(\chi^2_{\rm red}+\chi^2_{\rm blue})/{\rm d.o.f.}$, where $\chi^2_{\rm red}$ and $\chi^2_{\rm blue}$ are chi-square values for fits to the red and the blue galaxy correlation functions.
\sm{For the chi-square value for each color, we calculate
\begin{equation}
  \chi^2_{\rm color}=\sum_i\frac{(w_{p,{\rm obs},i}-w_{p,{\rm model},i})^2}{\sigma_{{\rm obs},i}^2+\sigma_{{\rm model},i}^2}.
  \label{eq:chi2}
\end{equation}
For the faintest bin, $-20<M_r<-19$, we do not include the three innermost bins for the $\chi^2$ value calculation since the SHAM underestimates the correlation function of the full luminosity sample in the small scales (see the top panel of Figure~\ref{wp_hod_lum}).}

\begin{figure}
  \includegraphics[width=8.5cm]{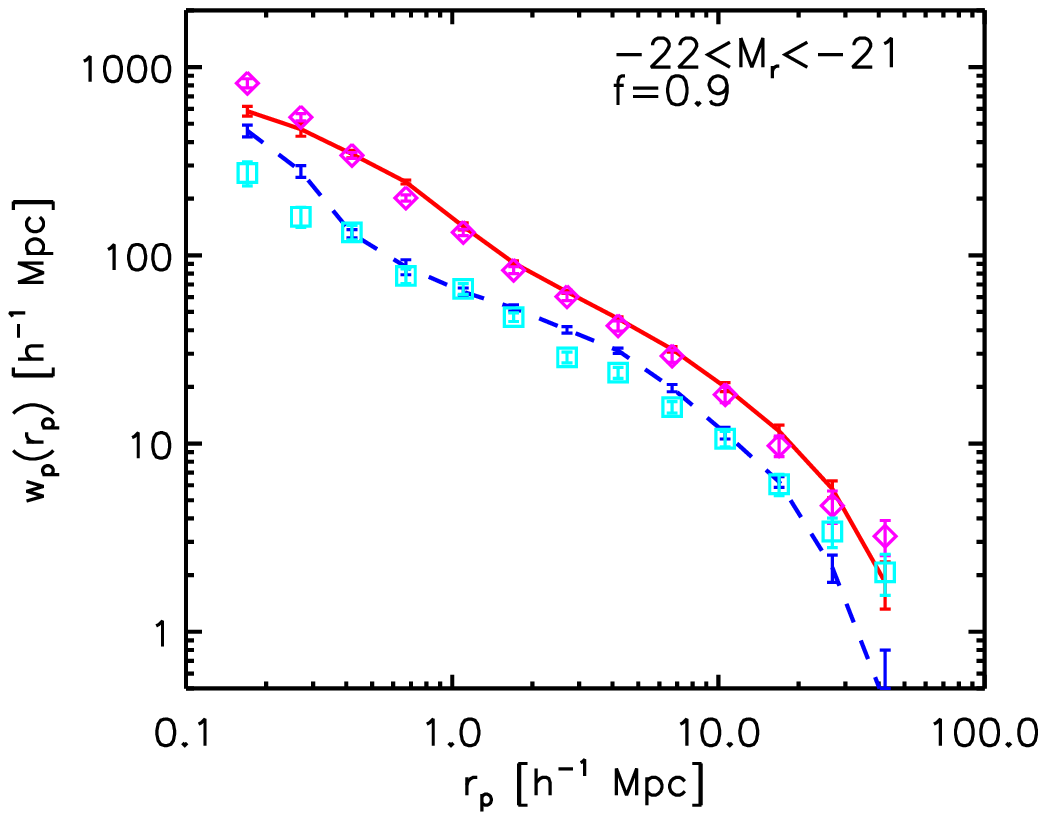}
  \includegraphics[width=8.5cm]{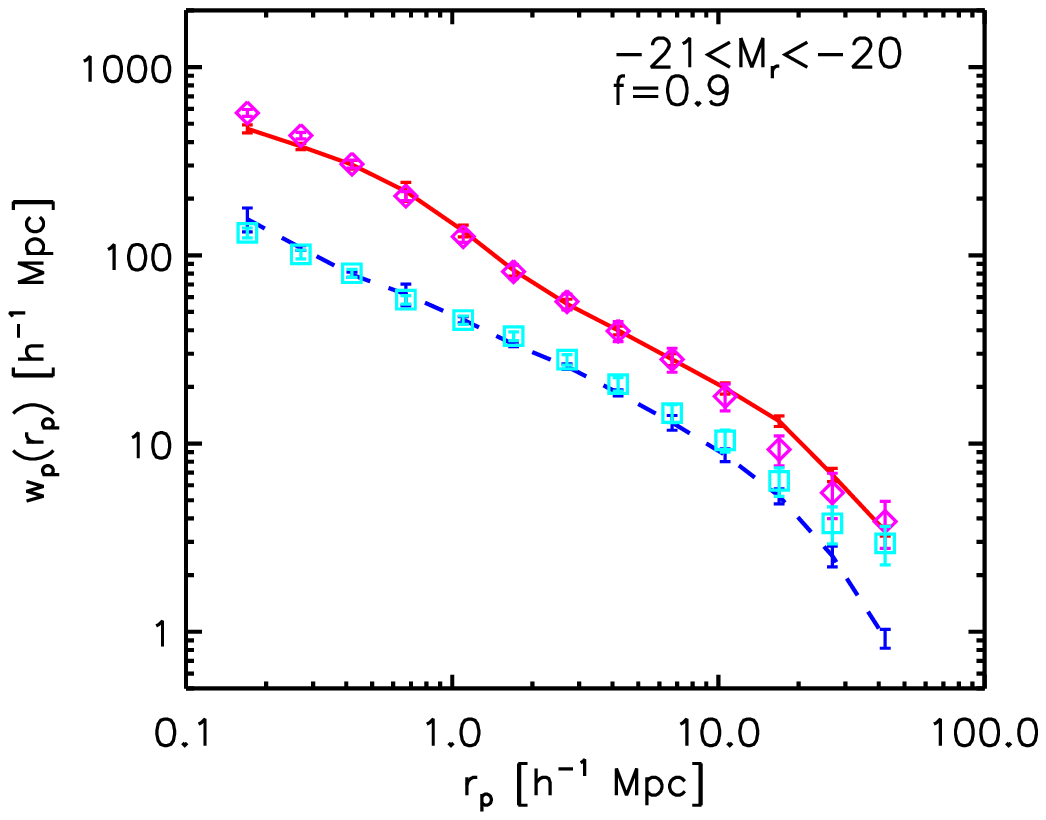}
  \includegraphics[width=8.5cm]{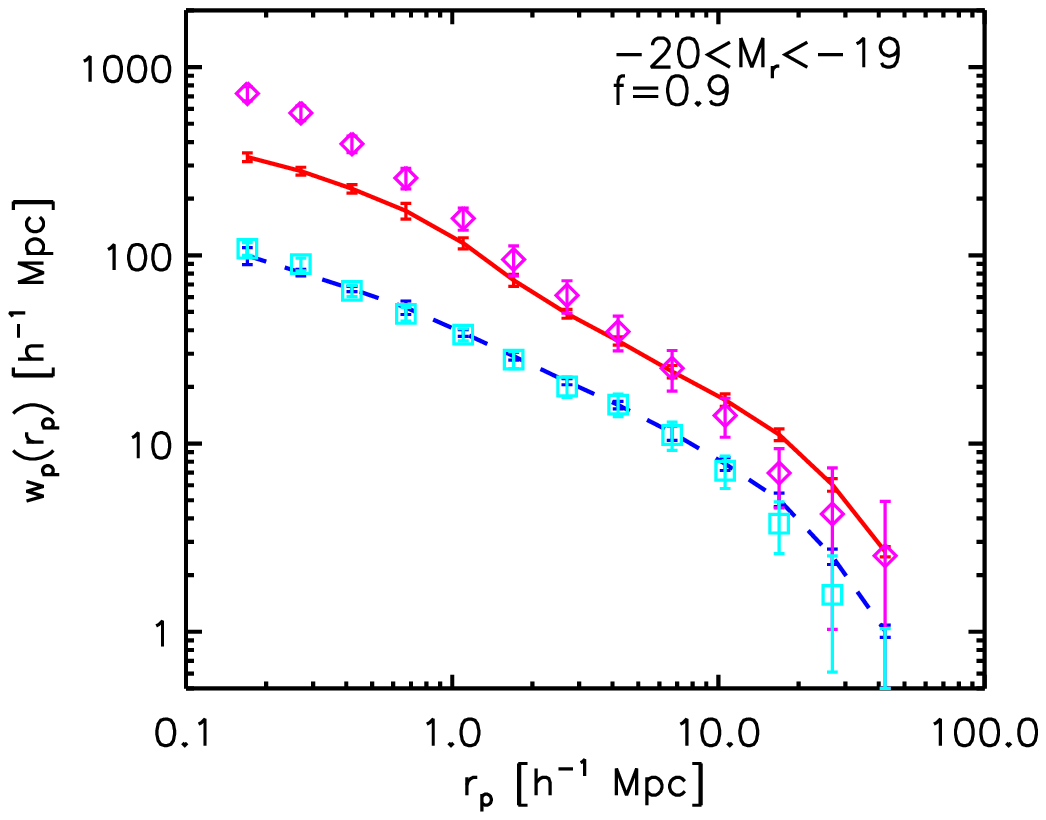}
  \caption{Comparison of the projected correlation functions between the best models in the age model (\fig{red solid and blue dashed lines}) and the observation \citep[\fig{open diamonds and boxes with error bars},][]{Zehavi11} for the three magnitude binned samples.}
  \label{wp_age}
\end{figure}

In the age model, we consider \sm{four values of $f=0.6, ~0.7, ~0.8$ and $0.9$.}
\sm{For the all-galaxy sample, we find that $f=0.9$ gives the least $\chi^2$ (see Figure~\ref{chi2} for the parameter dependence of $\chi^2$).
The best model results for the three samples are shown 
in Figure \ref{wp_age}.}
\fig{The red solid and the blue dashed} lines with error bars represent the predicted $w_p$ for the red and the blue galaxies, respectively.
The error bars are \sm{calculated by jack-knife sampling} with eight subvolumes.
\fig{The magenta diamonds and cyan boxes with error bars show the observational results for red and blue galaxies taken from \cite{Zehavi11}.}

As shown in Figure \ref{wp_age}, our age model reproduces the observed color-dependent clusterings very well.
The value of $f=0.9$ is relatively high but not surprising since the time evolution of the maximum circular velocity is slower than mass evolution.
For the Milky-Way sized distinct halo, $z_{\rm form}$ with $f=0.9$ corresponds to $z\simeq1-2$ on average \citep{Boylan-Kolchin10}.
In Table \ref{ave_zform}, we list the average values of $z_{\rm form}$ in the best models for each magnitude bin and color.
\begin{table}
  \begin{tabular}{@{}lccc}
    \hline
    ~ & bright & intermediate & faint \\
    \hline
    red & 1.77 & 2.44 & 3.04 \\
    blue & 0.52 & 0.78 & 0.98 \\
    \hline
  \end{tabular}
  \caption{The average values of $z_{\rm form}$ as a function of magnitude and color in the best fit age models, i.e., $f=0.9$ for all samples.}
  \label{ave_zform}
\end{table}

\begin{figure}
  \includegraphics[width=8.5cm]{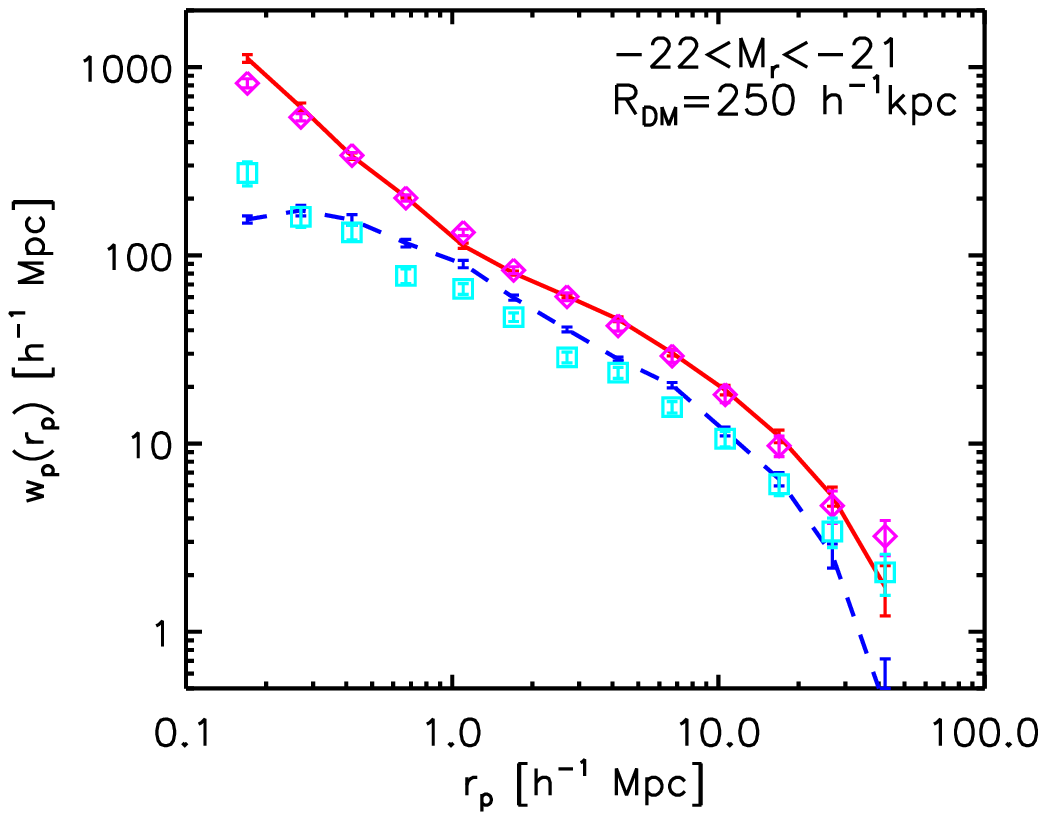}
  \includegraphics[width=8.5cm]{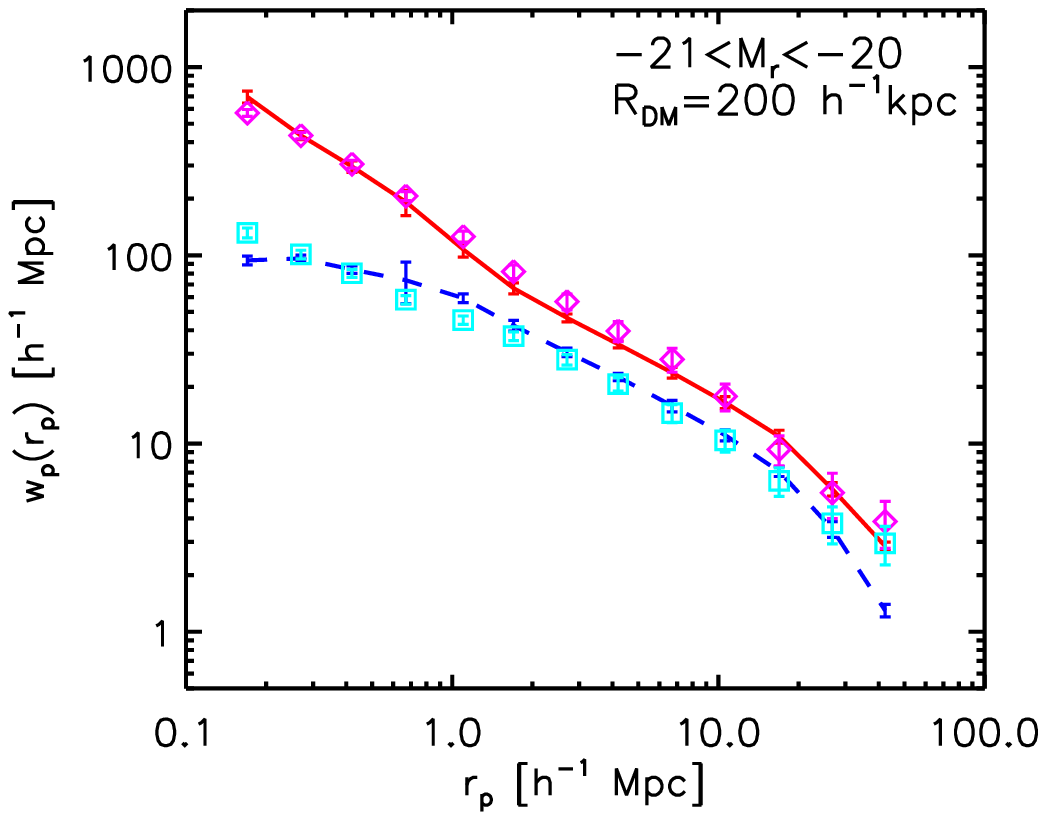}
  \includegraphics[width=8.5cm]{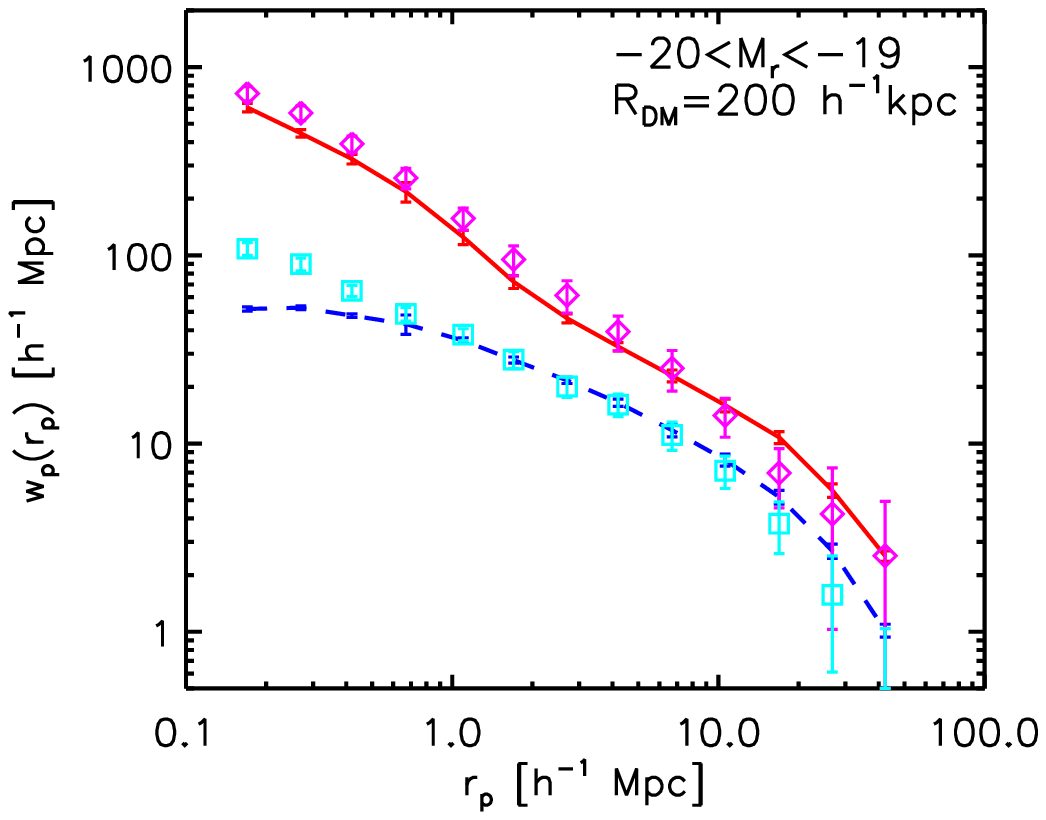}
  \caption{Same as Figure \ref{wp_age} but for the local density model.}
  \label{wp_loc}
\end{figure}

Next we present the results from the local density model.
We take \sm{five sphere sizes of $R_{\rm DM}=100,~200,~250,~300$ and $400~\hkpc$.}
We compare the best model results 
with the observation in Figure \ref{wp_loc}.
For the fainter and intermediate samples, we find that the model with $R_{\rm DM}=200~\hkpc$ is the best model.
Only for the brightest sample, the best fit is given by the model with $R_{\rm DM}=250~\hkpc$ \sm{(see Figure~\ref{chi2}).}

\sm{Figure~\ref{chi2} shows the $\chi^2$ values as a function of the parameters of the age and local density models, for each luminosity bin.
\begin{figure}
  \includegraphics[width=8.5cm]{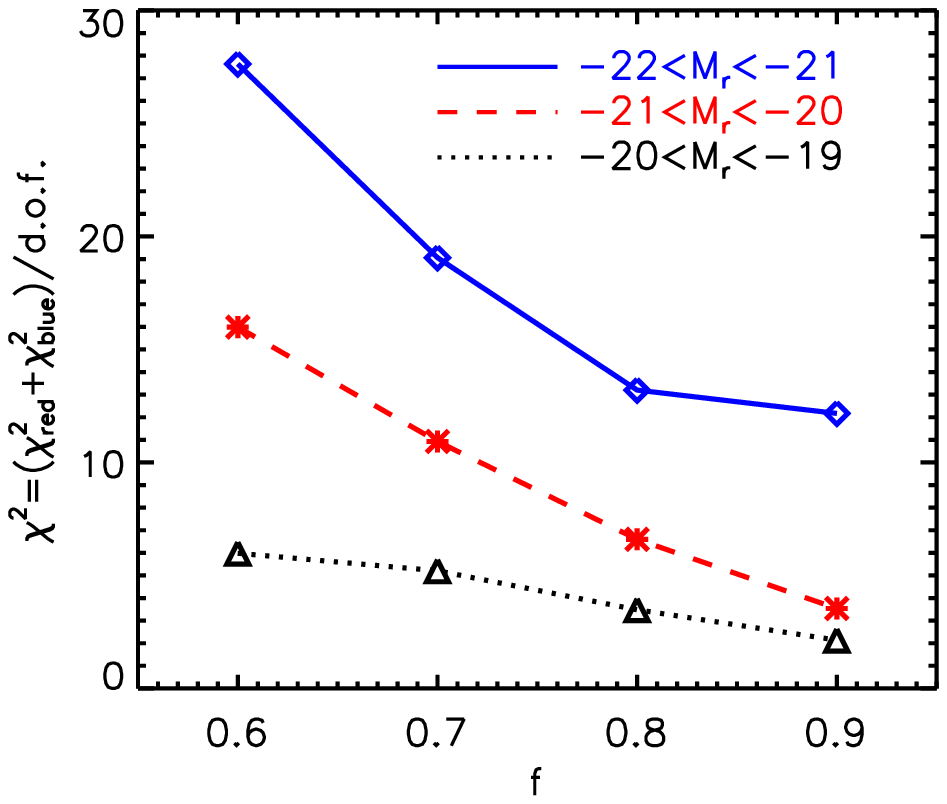}
  \includegraphics[width=8.5cm]{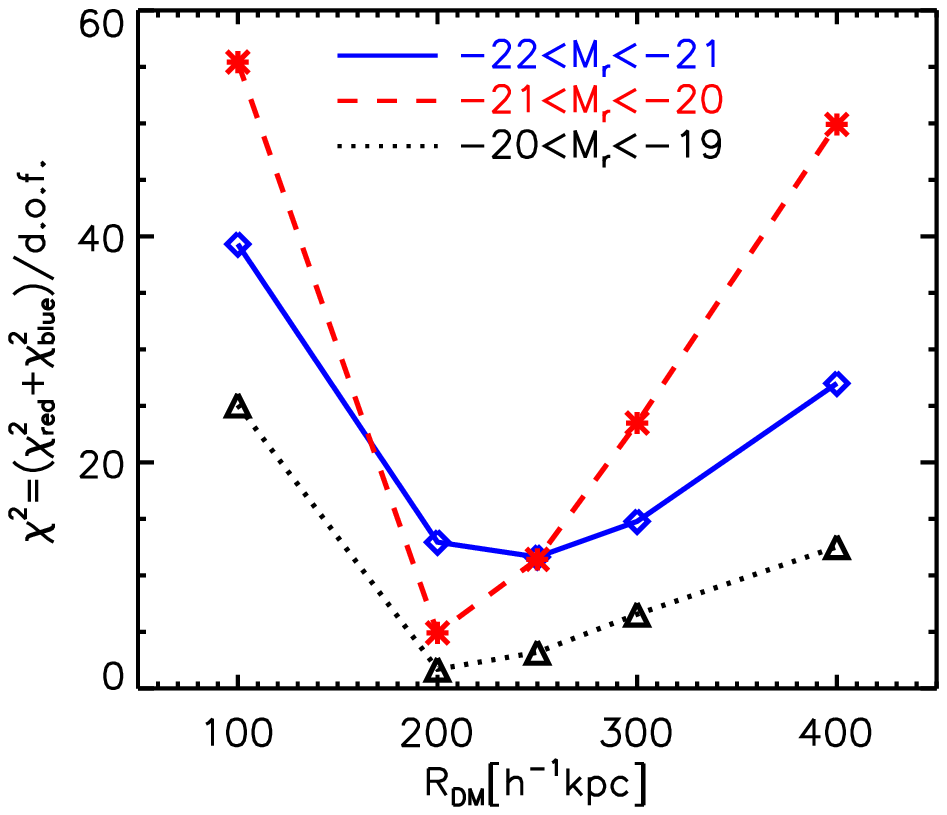}
  \caption{{\em Top}: The $\chi^2$ values as a function of the age model parameter $f$ for each luminosity bin. {\em Bottom}: The $\chi^2$ values as a function of the local density model parameter $R_{\rm DM}$ for each luminosity bin. Note that the $y$-axis scales are different among the two panels.}
  \label{chi2}
\end{figure}
In the top panel, it is shown that the $\chi^2$ values in the age model for all samples decrease monotonically as the model parameter $f$ increases.
The $\chi^2$ values of the brightest sample are relatively higher than those of other two samples.
The bottom panel shows that the $\chi^2$ values in the local density model for all samples have the minimum value around $R_{\rm DM}\simeq200~\hkpc$, and have higher values at smaller and larger radii.}

As well as the age model, our local density model shows good agreement with the observational results.
It should be noted that the local density model is motivated by the observed lensing profiles.
However the local density model systematically does not have enough clustering amplitude at the smallest scale, $r_p\la20~\hkpc$, for blue galaxies.
It is because that the local density model populates less massive subhalos with more blue central galaxies than the age model does.
We discuss this point in more details below.

\subsection{HODs from our models}
\label{sec:hod}
\begin{figure}
  \includegraphics[width=8.5cm]{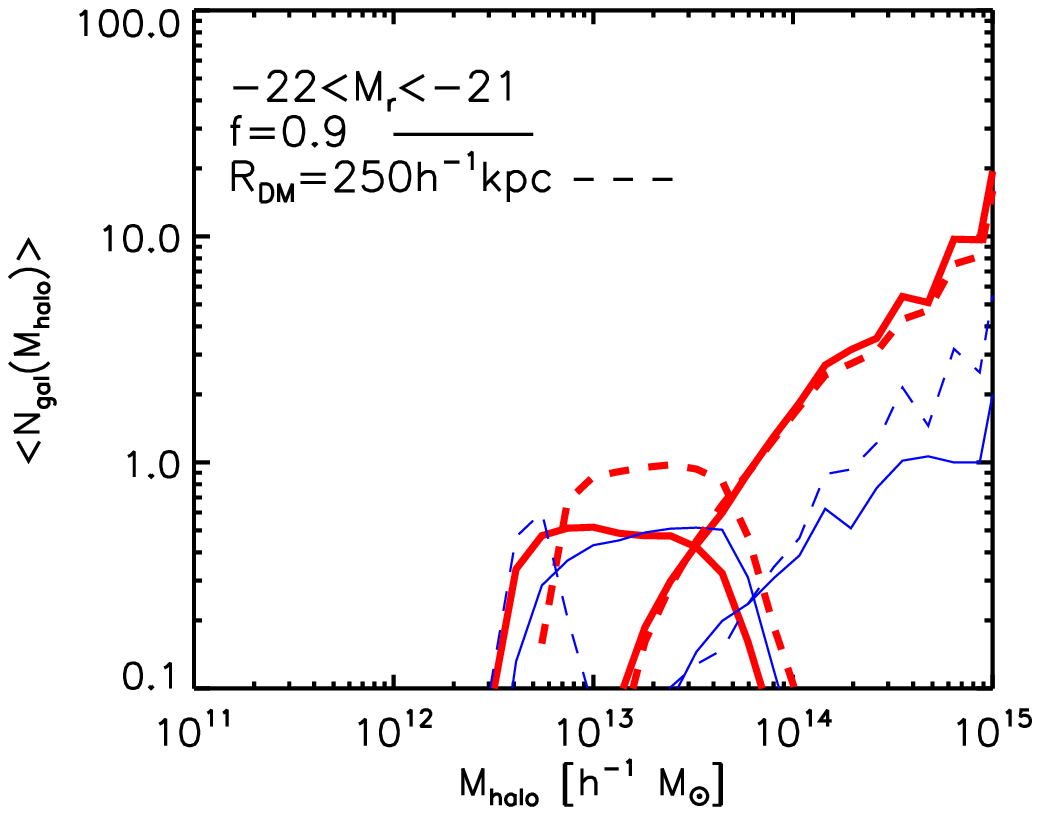}
  \includegraphics[width=8.5cm]{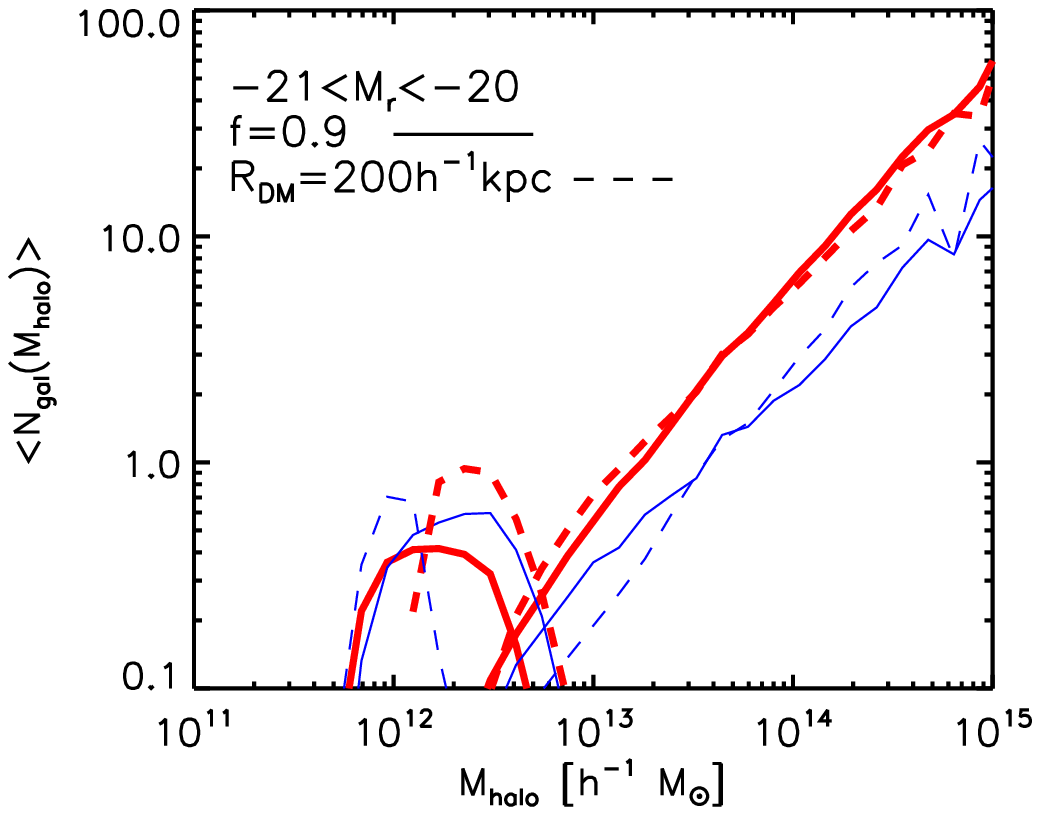}
  \includegraphics[width=8.5cm]{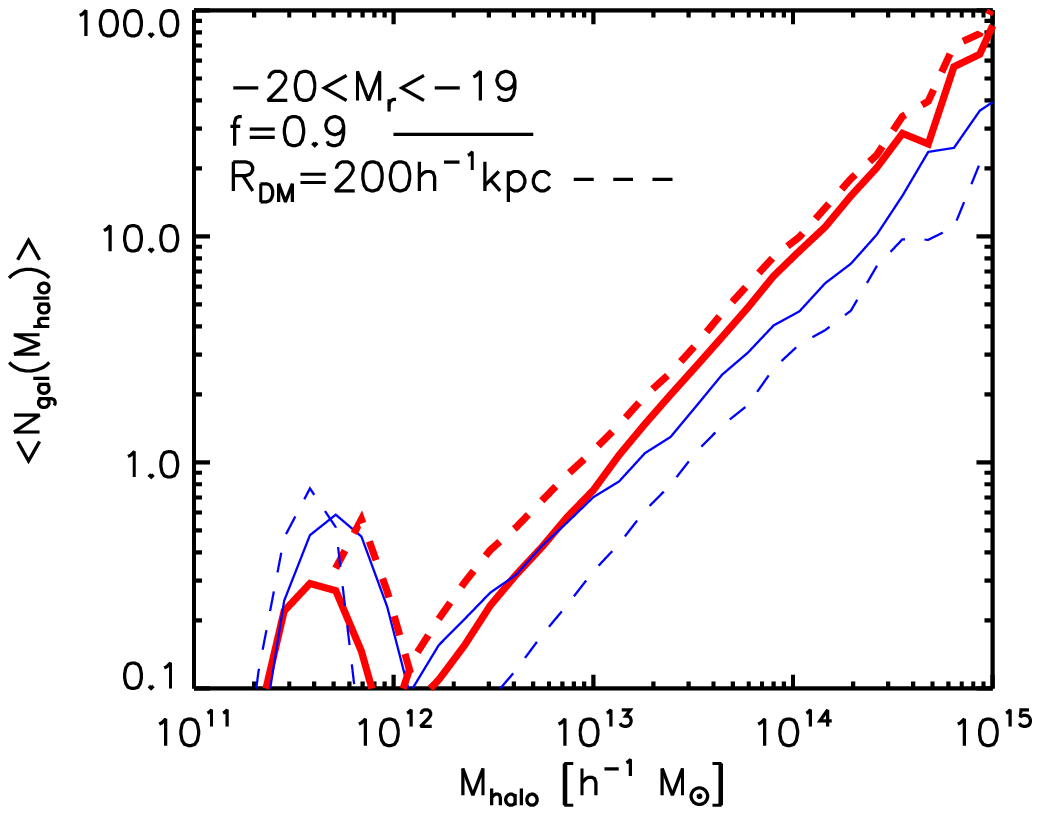}
  \caption{The obtained HODs from our two models for red and blue galaxies in the three magnitude bins. The quoted distinct halo mass is the FoF mass. The results from the age model and the local density model are shown by the solid and dashed lines, respectively. The thick red and thin blue lines represent the red and blue galaxy HODs.}
  \label{hod_col}
\end{figure}

The HODs for each magnitude bin obtained from the best models in our two models are shown in Figure \ref{hod_col}.
The results from the age model and the local density model are shown by the solid and dashed lines.
The thick red and thin blue lines represent the red and blue galaxy HODs, respectively.
Note that the quoted distinct halo mass in the horizontal axis is the FoF mass.
It can be seen that the shapes and amplitudes of HODs of satellite galaxies are very similar among our two models for red and blue galaxies.
The primary difference is found in HODs of central galaxies.
By construction, the host halo mass range of central is same in the two models.
In the local density model, central blue galaxies are hosted by low mass distinct halos which have few satellite ones.

We here compare our two models.
The most apparent difference between them is the clustering amplitude of blue galaxies at the smallest scale.
Systematic suppression of clustering amplitude can be found in the local density model.
At such scale, clustering is dominated by signals from central-satellite pairs.
As shown in Figure \ref{hod_col}, the local density model assigns the central blue galaxies to lower mass halos than the age model does whereas satellite galaxies live in more massive halos.
It means that the number of distinct halos with a central blue galaxy and more than one satellite blue satellites is decreased.
This suppresses the clustering amplitude at the small scale.

\cite{Zehavi11} calculated correlation functions analytically with the model HODs.
Their assumed HODs of red and blue galaxies have the same form but different amplitudes.
They also obtained very good agreement for both red and blue samples.
However our two models do not agree exactly with their HOD forms though both reproduce the observed correlation functions well.

\subsection{Color dependence of lensing profile}
\label{del_col}
We have shown that both the age and the local density models can be tuned to reproduce the observed color dependence of projected correlation functions reasonably well.
To discuss which model is better to assign galaxy color, we here consider the galaxy-galaxy lensing measurement $\Delta\Sigma$ as
\begin{eqnarray}
  \Delta\Sigma(r_p)=\gamma_{\rm t}(r_p)\Sigma_{\rm cr}=\bar\Sigma(<r_p)-\Sigma(r_p),
\end{eqnarray}
where $\Sigma$ is the surface mass density, $\gamma_{\rm t}$ is the tangential shear and $\Sigma_{\rm cr}=(c^2/4\pi G)(D_{\rm s}/D_{\rm ls}D_{\rm l})$ is the critical surface density \citep[see][for review]{BS01}. 
We project all simulation particles around each subhalo along a direction in the simulation box to measure the surface mass density \citep{Mandelbaum05,HayashiWhite08,Masaki12,NeisteinKhochfar12}.
Using the red/blue magnitude-binned subhalo catalogs obtained from our models, we calculate the mean profile for each catalog.

We compare the model predictions for red and blue galaxies with measurements for early and late type galaxies by \cite{Mandelbaum06}.
They used flux-limited samples and the average value of frac\_deV in $g,~r$ and $i$ bands to classify the early and late type galaxies ($\ge0.5$ for early types and $<0.5$ for late types).
The parameter frac\_deV specifies the galaxy flux profile as ${\rm frac\_deV}\times$(de Vaucouleurs profile)$+(1-{\rm frac\_deV})\times$(exponential profile).
As the {\it zeroth order approximation}, we regard early (late) type galaxies as red (blue) ones.
Using the SDSS DR7 \citep{Abazajian09}, we checked that 74\% of early (89\% of late) type galaxies actually lie in the red (blue) sequence for $-21<M_r<-20$.
For the fainter (brighter) sample, these rates are 78 (74) \% and 85 (91) \% for early-red and late-blue correspondences.
Therefore the early (late) type samples primarily consist of red (blue) galaxies.
Again, the red/blue galaxies are separated via Equation (\ref{divider}).
\cite{Sheldon04} showed that early (late) types and red (blue) galaxies have very similar lensing profiles but with the different color division and early/late type definition.

\begin{figure}
  \includegraphics[width=8.5cm]{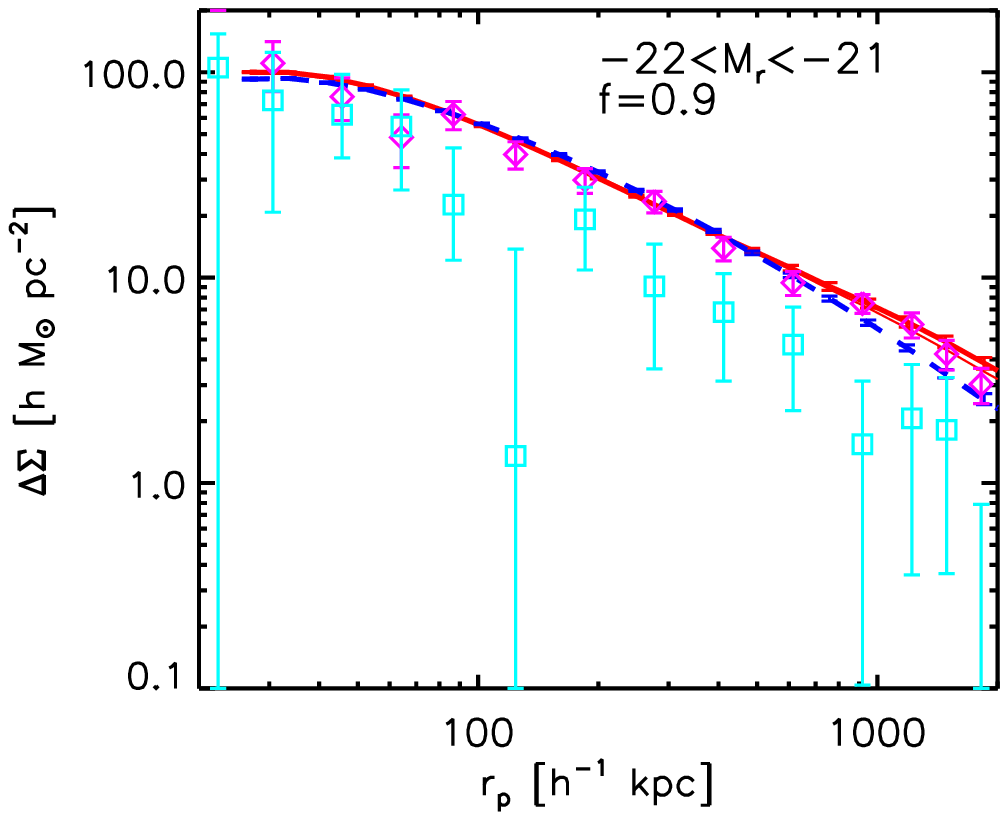}
  \includegraphics[width=8.5cm]{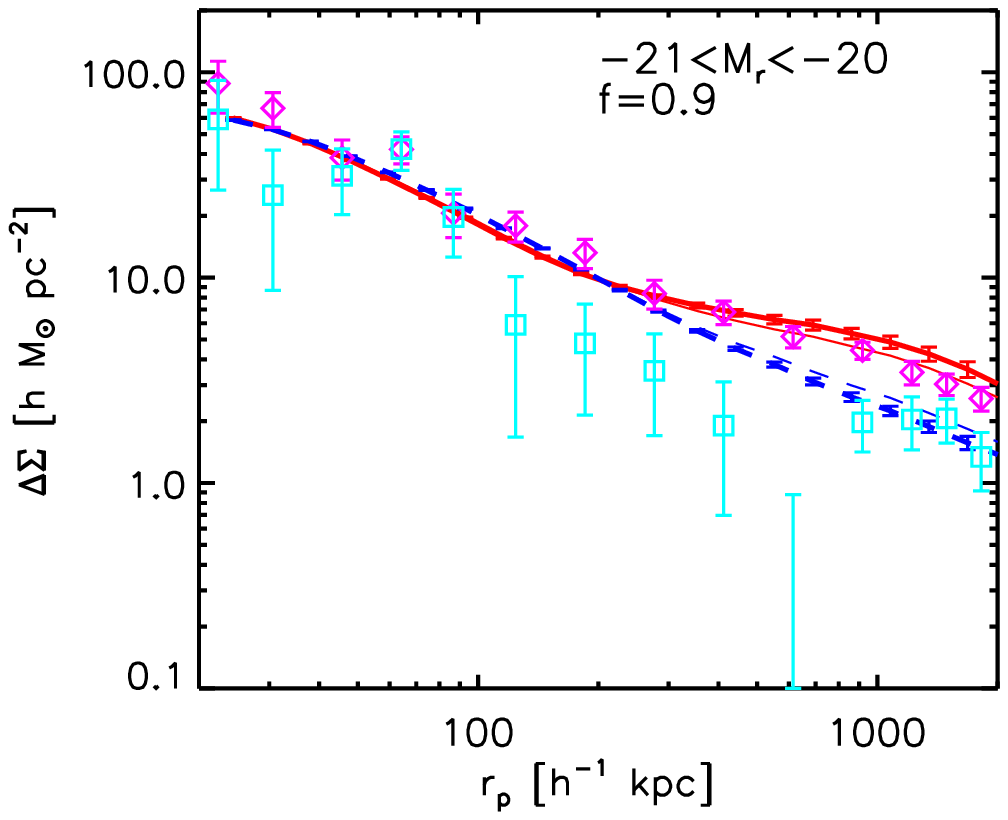}
  \includegraphics[width=8.5cm]{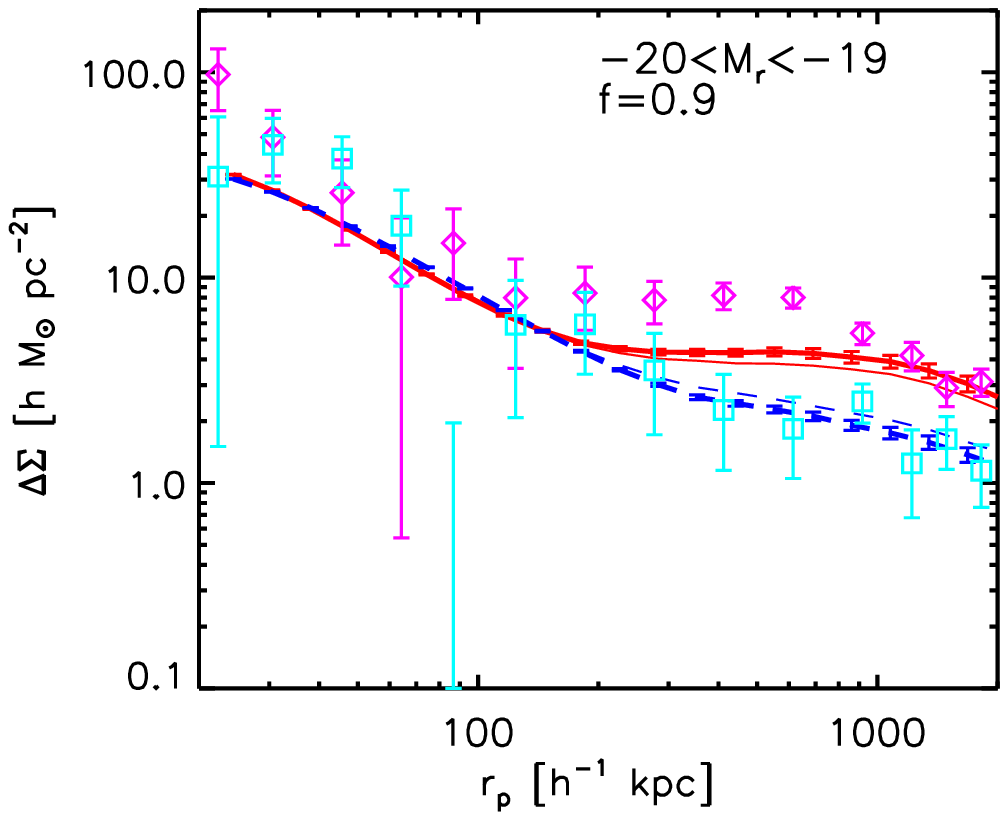}
  \caption{Surface mass density profile around galaxies. \fig{The red solid and blue dashed thick lines} with error bars represent the predictions for the red and blue galaxies in the age model. \fig{The magenta diamonds and cyan boxes} with error bars are the observational results from the SDSS \citep{Mandelbaum06} and show the early and late type galaxies' mass profile, respectively. The thin lines show results for the simulated early/late type samples (see text).}
  \label{del_age}
\end{figure}

In Figure \ref{del_age}, we show the results from our age model by thick lines with error bars for the three luminosity binned samples.
\fig{The red solid and blue dashed lines} are results for red and blue galaxies, respectively.
The error bars are taken from twenty-seven subsamples of subhalos.
\fig{The magenta diamonds and the cyan boxes} with error bars are the lensing profiles of early and late type galaxies measured by \cite{Mandelbaum06}.
For the sample of $-22<M_r<-21$, we simply show the number weighted averages of the observed lensing profiles of $-22<M_r<-21.5$ and $-21.5<M_r<-21$ samples.
The used subhalo catalogs are from models with $f=0.9$ for the \sm{three magnitude binned} samples.
It can be seen that the age model predictions for the red galaxies agree with the observational results of the early type galaxies well in the three magnitude bins.
However the age model does not trace the observation of the late type galaxies of the intermediate and the brightest samples at $100~\hkpc\la r_p\la1~\hMpc$ at all.

\begin{figure}
  \includegraphics[width=8.5cm]{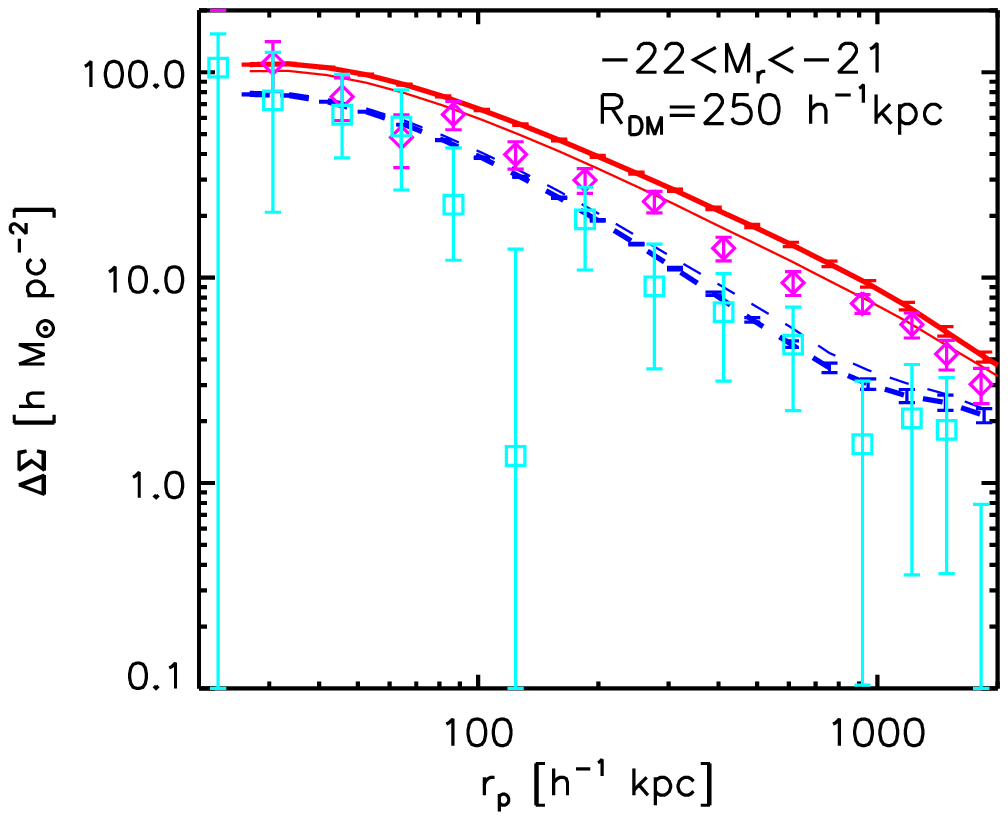}
  \includegraphics[width=8.5cm]{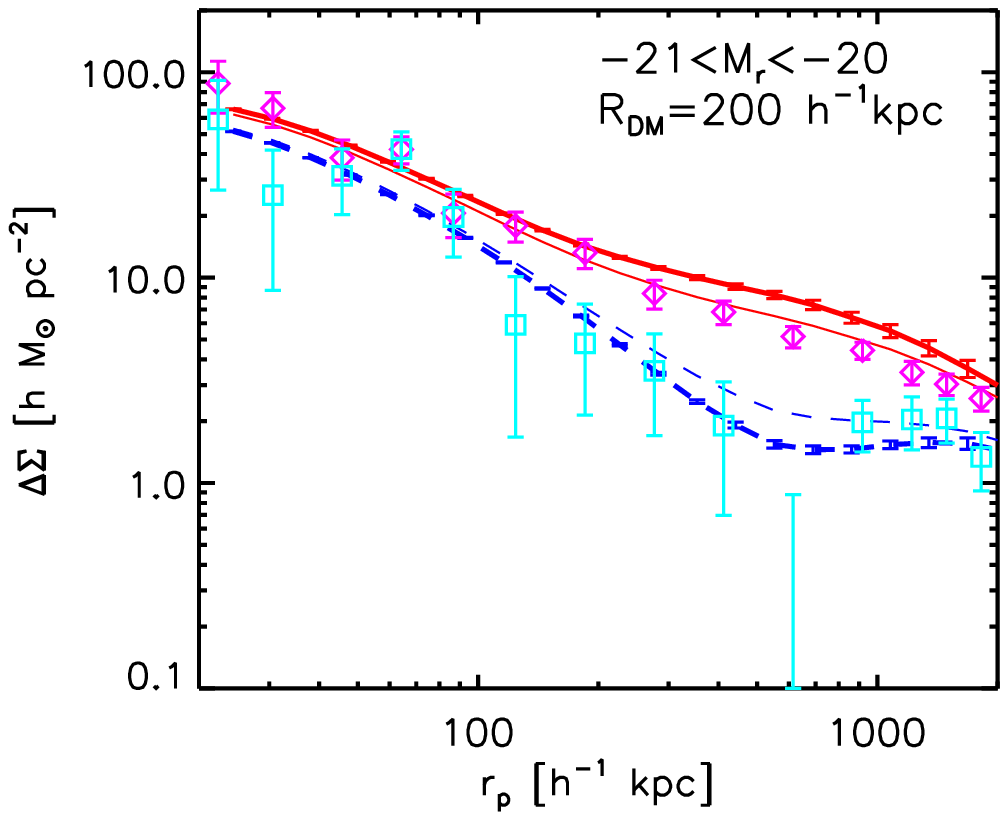}
  \includegraphics[width=8.5cm]{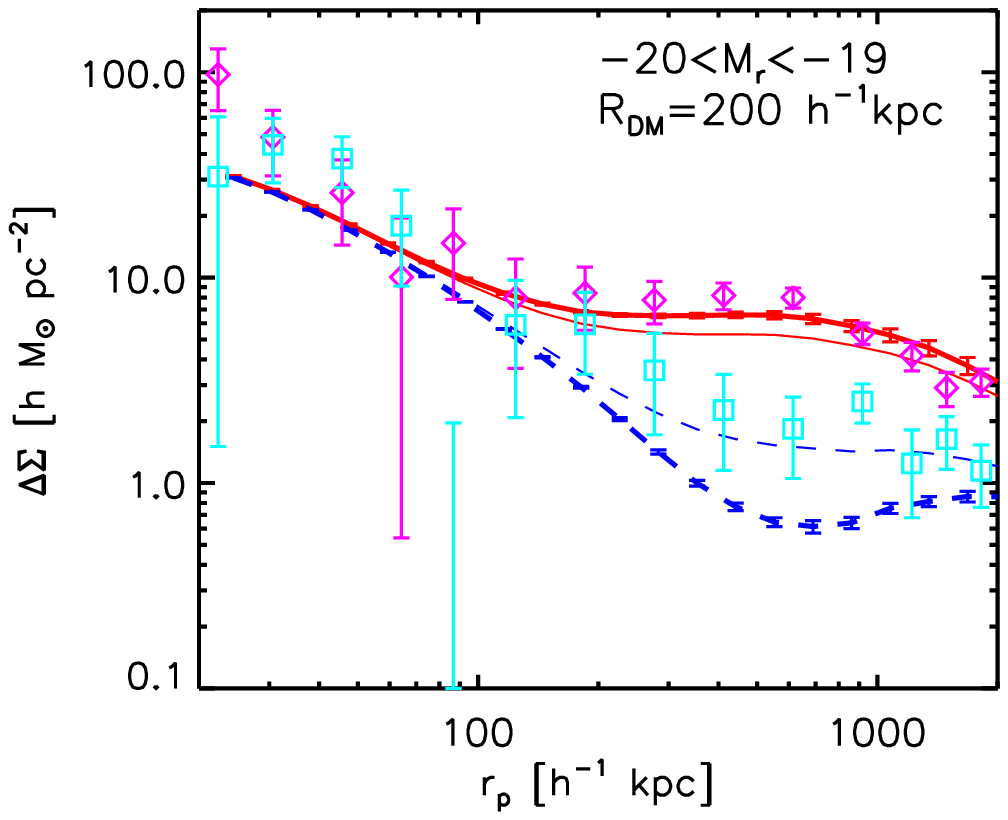}
  \caption{Same as Figure \ref{del_age} but for the local density model.}
  \label{del_loc}
\end{figure}

The results from our local density model are shown in Figure \ref{del_loc} in the same way as Figure \ref{del_age}.
For the intermediate and the faintest bins, the adopted size for local DM density measuring is $R_{\rm DM}=200~\hkpc$.
For the brightest bin, $R_{\rm DM}=250~\hkpc$ is adopted.
As well as the age model, agreements between the local density model and the observation are good.
The most striking difference from the age model predictions is larger color split at $r_p\ga300~\hkpc$.
This is because the blue galaxies in the local density model live in the less dense region than red ones do.
The split is large enough to trace the late type lensing profiles.

\begin{figure}
  \includegraphics[width=8.5cm]{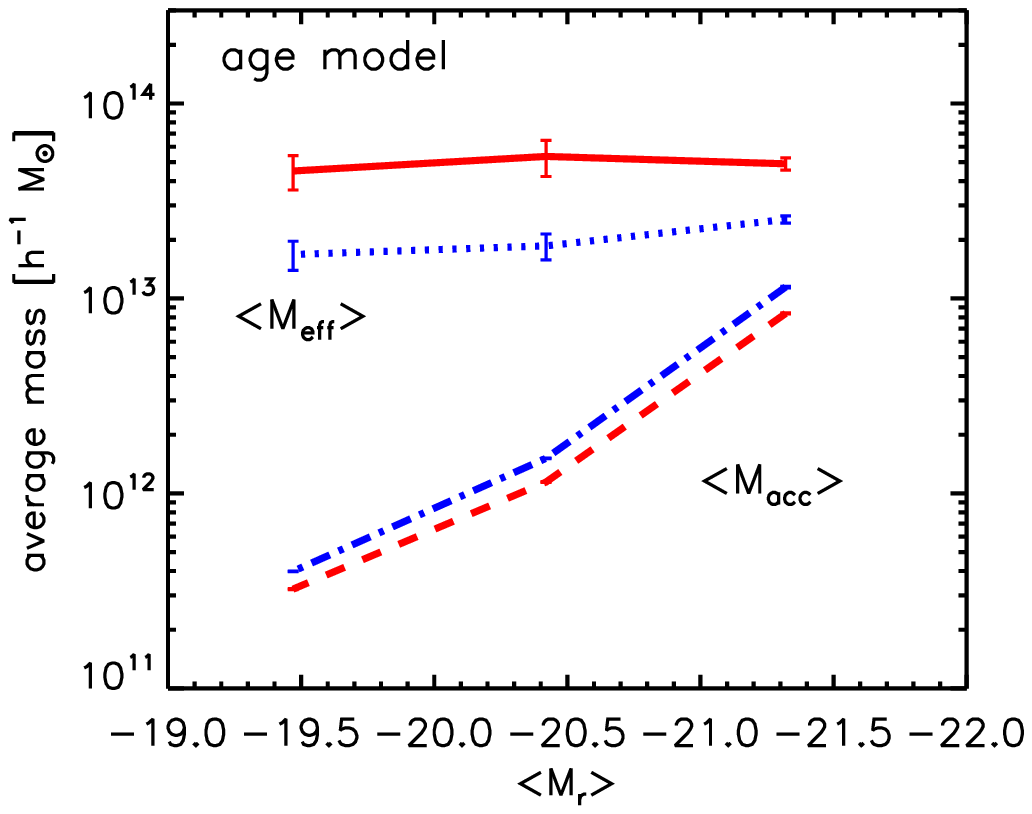}
  \includegraphics[width=8.5cm]{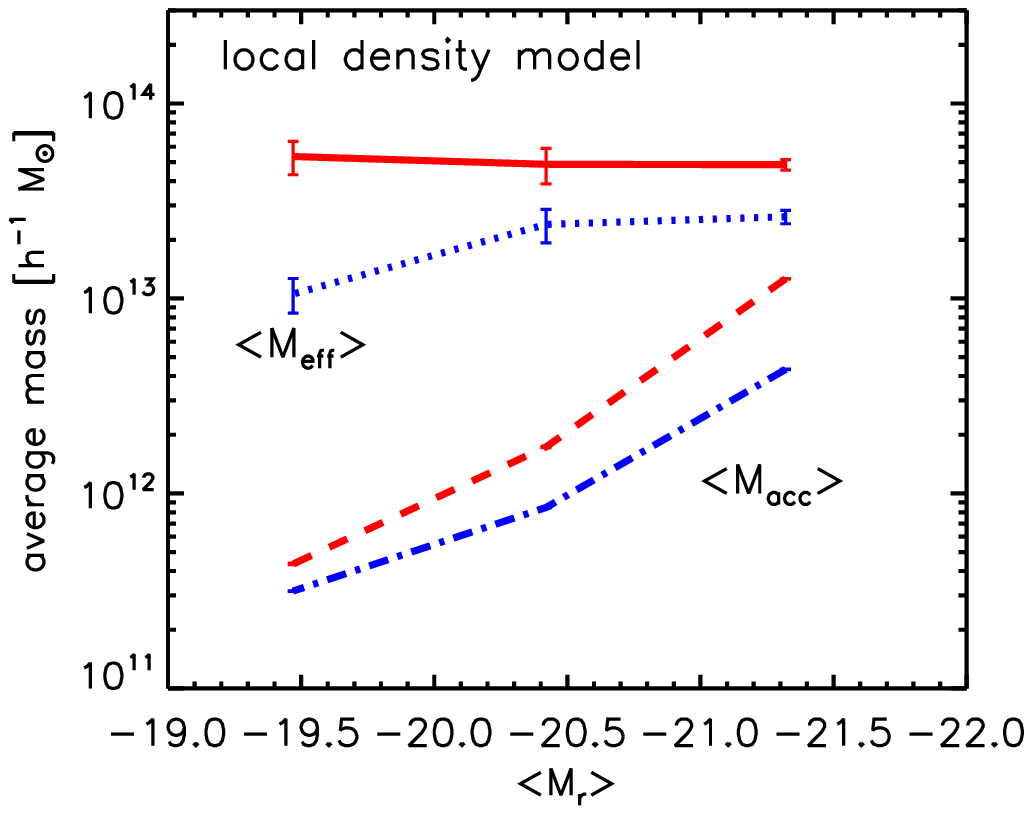}
  \caption{The effective mass of distinct host  halos $\langle M_{\rm eff}\rangle$　and the average mass of host subhalos $\langle M_{\rm acc}\rangle$ as a function of magnitude. $\langle M_{\rm eff}\rangle$ for red and blue galaxies are shown by solid and dotted lines, respectively. $\langle M_{\rm acc}\rangle$ for red and blue galaxies are shown by dashed and dotted-dashed lines, respectively. 
\sm{The error bars are calculated by jack-knife resampling of twenty seven subvolumes. The error bars for $\langle M_{\rm acc} \rangle$ are very small and thus are not visible in this plot.} {\em Top}: The results for the age model. {\em Bottom}: Same as Top panel but for the local density model.}
  \label{fig:ave_mass}
\end{figure}

We discuss the difference between our two models more quantitatively here.
We estimate the effective mass of host distinct halos $\langle M_{\rm eff}\rangle$ and the average mass of host subhalos $\langle M_{\rm acc}\rangle$ for each magnitude bin and color.
$\langle M_{\rm eff}\rangle$ is the HOD weighted average mass of host distinct halos $M_{\rm halo}$ and is calculated as
\begin{equation}
  \langle M_{\rm eff}\rangle=\int dM_{\rm halo} \frac{dn}{dM_{\rm halo}} M_{\rm halo}\frac{\langle N_{\rm gal}(M_{\rm halo})\rangle}{n_{\rm gal}},
\end{equation}
where $dn/dM_{\rm halo}$ is the distinct halo mass function.
The distinct halo mass $M_{\rm halo}$ comes from the FoF algorithm.
The \sm{characteristic} subhalo mass is determined by the {\it SubFind} algorithm at the accretion epoch.
We show the masses as a function of magnitude in Figure \ref{fig:ave_mass} for our two models.
The quoted magnitude is the average value for each magnitude bin which can be found in \cite{Zehavi11}.
The top and bottom panels show the results from the age and the local density models, respectively.
We see that $\langle M_{\rm eff}\rangle$ of red galaxies is always greater than that of blue galaxies in both the age and the local density models.
This is consistent with the fact that the model predicted lensing profiles of red galaxies have higher amplitude than those of blue galaxies at all scales.
Also in the local density model, $\langle M_{\rm acc}\rangle$ of red galaxies is greater than that of blue galaxies.
However in the age model, $\langle M_{\rm acc}\rangle$ of blue galaxies is more massive than that of red galaxies.
This trend can be found in HODs of central galaxies from the age model (see Figure \ref{hod_col}).
This is because more massive halos have more rapid mass growth on average \citep[see e.g.,][]{Wechsler02}.
Hence, with our definition of the subhalo formation epoch, more massive subhalos tend to be blue galaxies in our age model.
Thus the resulting low (high) mass of host subhalos of red (blue) galaxies in the age model would make the color split of the lensing profiles smaller than in the local density model even at $r_p\la {\rm a~few~}\times100~\hkpc$.

To discuss further, we \sm{attempt to make a ``fairer''} comparison with the observation of \cite{Mandelbaum06}.
We match the red/blue fraction of our simulated galaxies in a magnitude bin
to the fraction we obtain from the actual SDSS galaxy catalog for the early/late type sample.
To this end, we blend the red/blue galaxies randomly.

The results of lensing measurements from the matched early/late type galaxy catalogs are shown by thin lines in Figures \ref{del_age} and \ref{del_loc} without error bars.
\fig{The red solid and blue dashed thin lines} represent lensing profiles of the early and late types, respectively.
We see the effect of the color fraction matching clearly at $r_p>300~\hkpc$.
Figure \ref{del_loc} shows fairly good agreement between the predictions from the modified subhalo samples with the local density model and the observations.
However the agreement is less impressive for the age model subhalo samples, as can be seen in Figure \ref{del_age}.
This implies that the local DM density is more crucial for the galaxy color assignment model than the subhalo formation epoch defined via Equation (\ref{zform}).

\sm{\subsection{Scatter in the local DM density-color relation?}
\subsubsection{Small scale clustering}}
By comparing the observed color dependence of clustering and mass profile, our local DM density model can be regarded as the better model to assign galaxy color to DM subhalos.
However reproducing the clustering amplitude of the blue galaxies at $r_p\simeq10~\hkpc$ remains a challenge for our model.
It is due to that the model traces the DM halo mass too strongly (see Section \ref{sec:hod}).
It can be expected that introducing scatter in the local DM density-color relation has an impact on the small scale clustering amplitudes.
We here study how much the inclusion of scatter improves the agreement with the SDSS measurements.

We adopt a very simple approach to introduce scatter.
We perturb the local DM density $\rho_{\rm DM,m}$ measured in the sphere of $R_{\rm DM}$ as follows.
For each subhalo, logarithm of the perturbed local density $\rho_{\rm DM,p}$ is drawn from the Gaussian distribution with
\begin{eqnarray}
  {\rm average:}&~&\mu=\log_{10}(\rho_{\rm DM,m}),\\
  {\rm standard~deviation:}&~&\sigma=\sigma_{\rm p}\times\log_{10}(\rho_{\rm DM,m}).
\end{eqnarray}
For $\sigma_{\rm p}$, we take $0.1,~0.2,~0.3$ and $0.4$.
We adopt these four values for the six sphere sizes of $R_{\rm DM}=200,~300,~400,~500,~600$ and $750~\hkpc$, i.e., we have twenty-four combinations of $\sigma_{\rm p}$ and $R_{\rm DM}$.
We divide the luminosity binned subhalo catalog into the color binned subsamples according to the ranking of the perturbed local DM density $\rho_{\rm DM,p}$ rather than that of the originally measured density $\rho_{\rm DM,m}$.
Then we search for the best model among the twenty-four models to recover the observed color dependences of galaxy correlation functions.

We find that, for larger $\sigma_{\rm p}$, the split of the projected red and blue correlation functions is smaller.
This is just because introducing scatter is similar to averaging the original red and blue galaxy distributions in the no-scatter model.
Hence the best sphere size for the scatter model would be larger than what we obtained for the no-scatter model since the color split is larger for larger $R_{\rm DM}$ (see Figure \ref{wp_hod_loc}).

\begin{figure}
  \includegraphics[width=8.5cm]{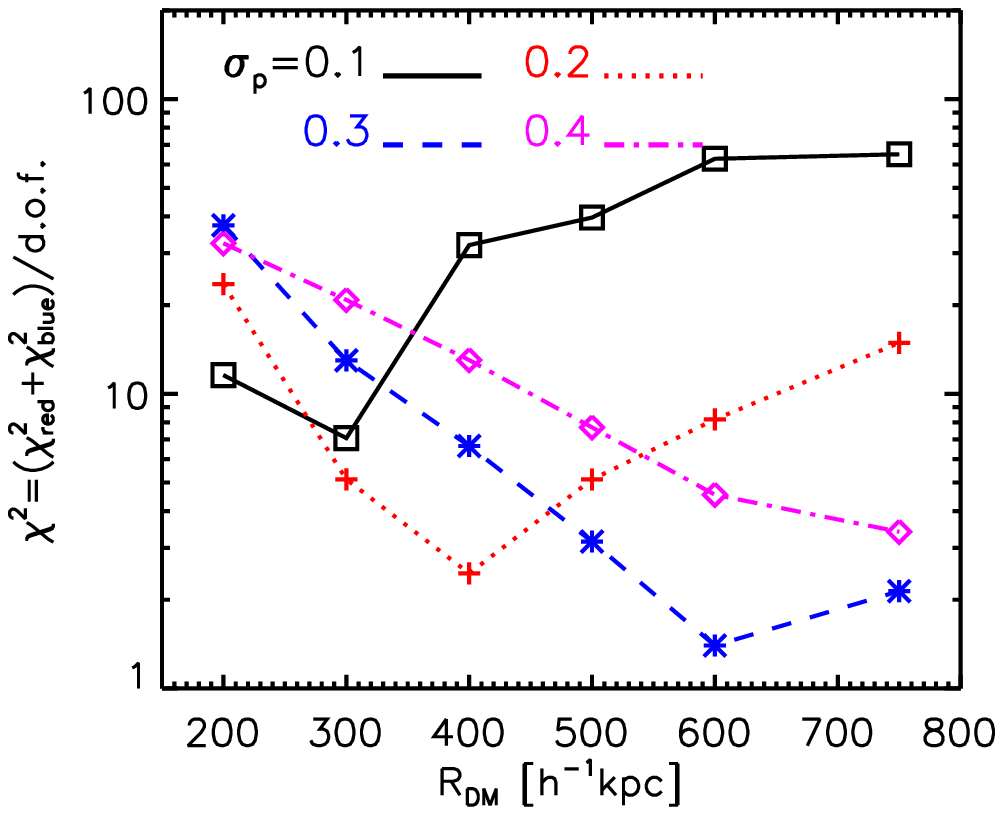}
  \includegraphics[width=8.5cm]{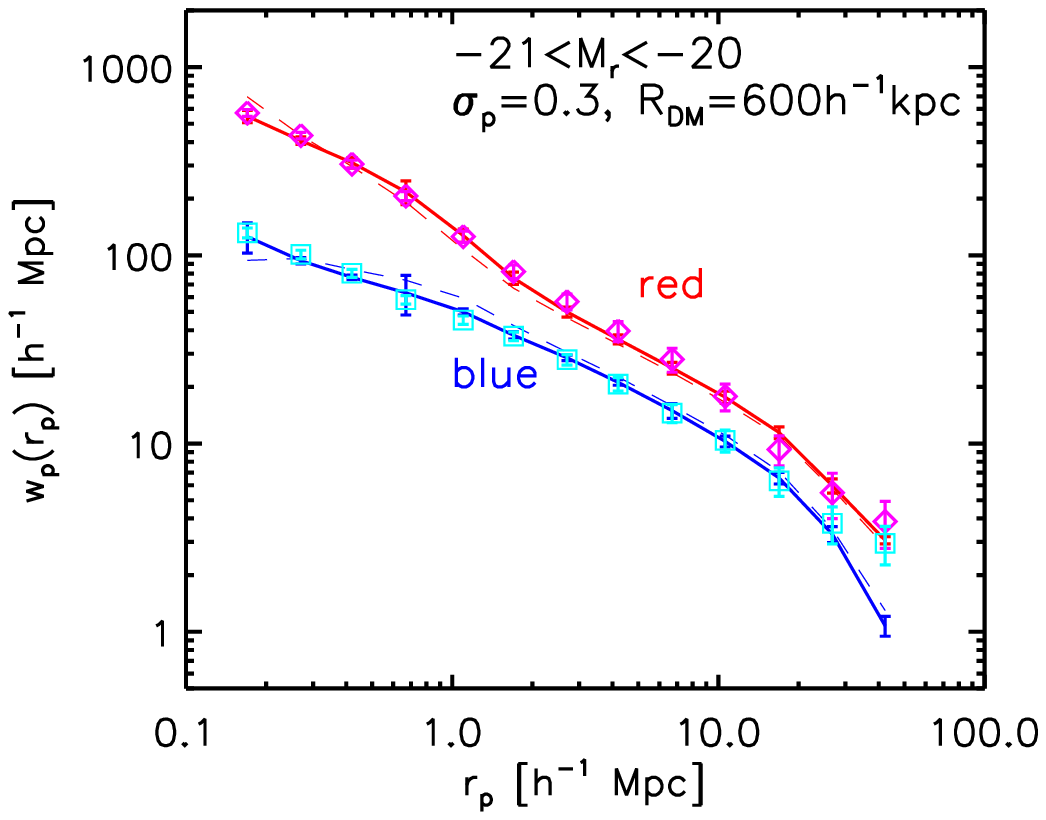}
  \includegraphics[width=8.5cm]{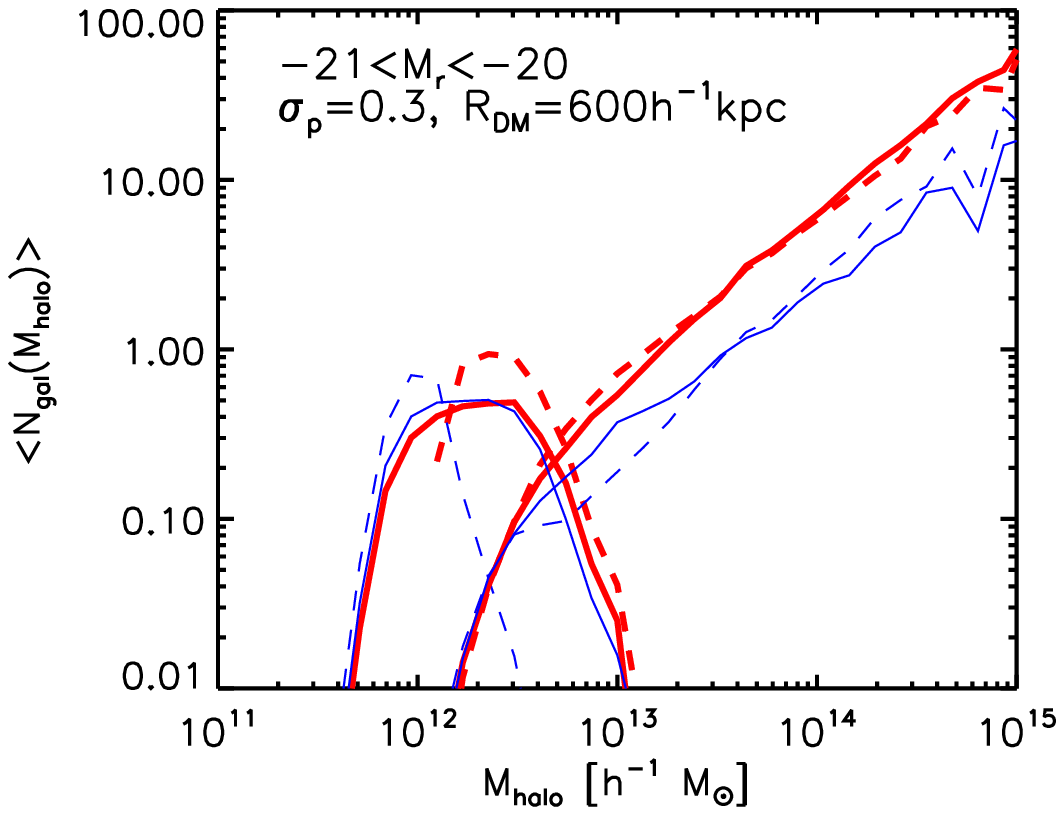}
  \caption{{\em Top}: The $\chi^2$ values as a function of $\sigma_{\rm p}$ and $R_{\rm DM}$ which we use to search the best model among our twenty-four combinations of $\sigma_{\rm p}$ and $R_{\rm DM}$. {\em Middle}: The projected correlation functions of red and blue galaxies from the best model of $\sigma_{\rm p}=0.3$ and $R_{\rm DM}=600~\hkpc$. For comparison, we also show the predictions from the local density model without scatter by the thin dashed lines. {\em Bottom}: \fig{The thick red and thin blue solid lines} show HODs of red and blue galaxies from the best model with scatter, respectively. The dashed lines show the results from the no-scatter model.}
  \label{wp_hod_sca}
\end{figure}

Here we study the impact of scatter on the red and blue galaxy correlation functions of the intermediate magnitude sample of $-21<M_r<-20$.
The top panel of Figure \ref{wp_hod_sca} shows the $\chi^2$ values for the projected correlation functions of red and blue galaxies as a function of $\sigma_{\rm p}$ and $R_{\rm DM}$.
From the panel, we find that a combination of 
\begin{eqnarray}
  R_{\rm DM}=600~\hkpc~~~{\rm and}~~~\sigma_{\rm p}=0.3
\end{eqnarray}
gives the least $\chi^2$ value among the twenty-four models.
The minimum $\chi^2$ value is $1.4$ and smaller than that among the no-scatter model of $4.9$.
In the middle panel of Figure \ref{wp_hod_sca}, we show the predicted color dependences of projected correlation function.
For comparison, we also show the results from the ``non-perturbed'' local density model of $R_{\rm DM}=200~\hkpc$ by thin dashed lines.
The suppression of blue galaxy clustering in the smallest scale is no longer seen and the shapes of red and blue galaxy correlation functions are very similar to those of the SDSS measurements.
It is because that the blue central galaxies are assigned into more massive distinct halos than in the no-scatter model, as can be clearly seen in the HODs shown by the solid lines in the bottom panel.
As in the middle panel, the dashed lines represent the HODs from the non-perturbed local density model.
The panel shows that the inclusion of scatter alters HODs of central galaxies.
It should be noted that agreements of the no-scatter and scatter models with the SDSS results are both acceptable although improvements by introducing scatter is impressive.

\sm{\subsubsection{Finer color bin samples}}
To make the subhalo property-galaxy color relation more concrete, it is crucial to consider measurements for finer color bin samples.
It is also natural to expect that scatter in the local DM density-color relation becomes more important for the correlation functions in finer color bins.
In this section we study the impact of the inclusion of scatter on the projected correlation functions for finer color bins.
\cite{Zehavi11} divided the faintest magnitude bin sample in this paper ($-20<M_r<-19$) into the finer color bins (reddest, red sequence, redder, green, bluer and bluest) and measured the projected correlation functions for each bins (see their Figures 18 and 19).
They found that redder galaxies have higher clustering amplitudes at the overall scales of $0.1<r_p<10~\hMpc$.
We here make finer color bins, that is, reddest, redder, bluer and bluest samples following \cite{Zehavi11}.
Similar to the way we create the red and blue galaxies, we divide the faintest magnitude bin sample into four fine color bins according to the local DM density and match each number density of subhalo subsample to that of SDSS galaxy sample.
We use not only the originally measured local DM density but also the perturbed density as the galaxy color proxy.

\begin{figure}
  \includegraphics[width=8.5cm]{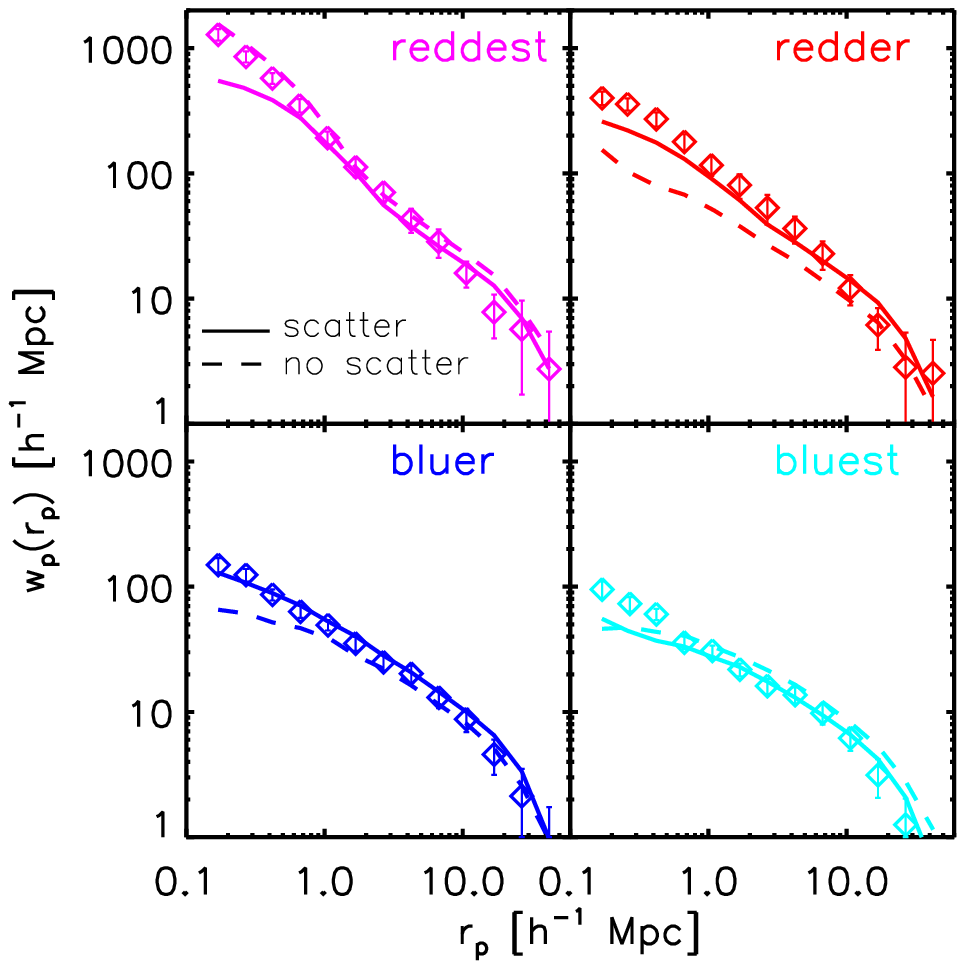}
  \includegraphics[width=8.5cm]{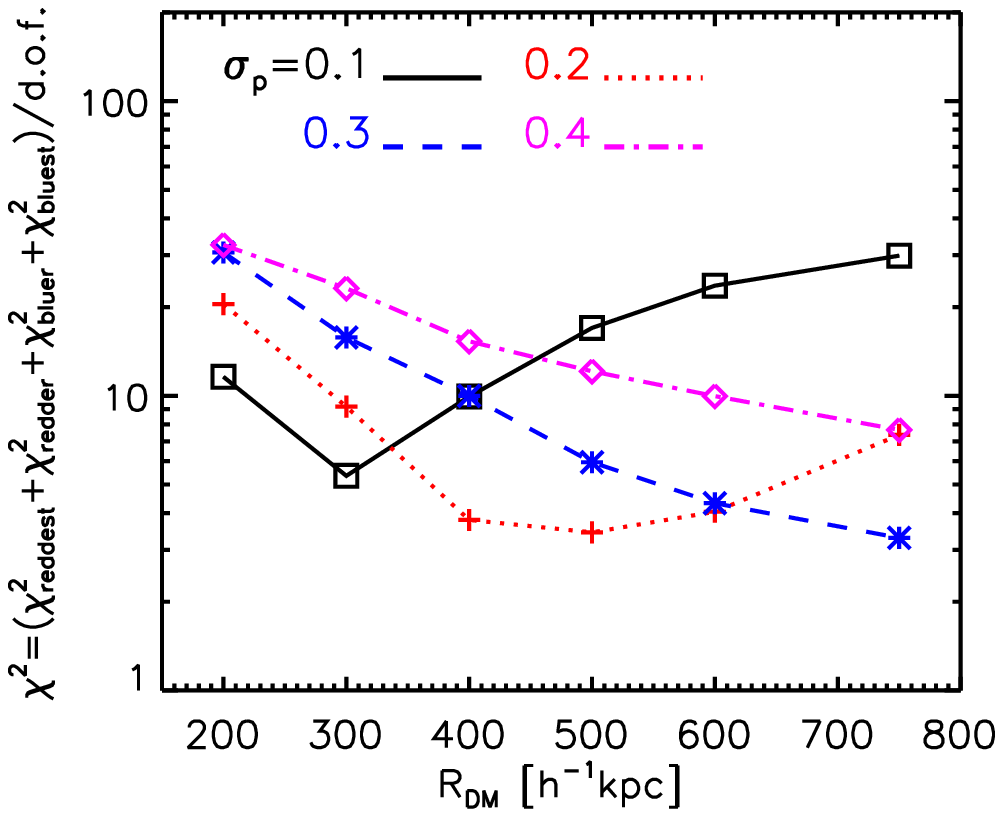}
  \caption{{\em Top}: The projected correlation functions in the finer color bins of the faintest, i.e., $-20<M_r<-19$, sample. The data points are from \citet{Zehavi11}. The dashed and solid lines are the results from the best no-scatter and scatter local density models, respectively. For the no-scatter model, we use $R_{\rm DM}=200~\hkpc$. We adopt the combination of $\sigma_{\rm p}=0.3$ and $R_{\rm DM}=750~\hkpc$ for the scatter model. {\em Bottom}: The chi-square values for the scatter model as a function of $\sigma_{\rm p}$ and $R_{\rm DM}$.}
  \label{wp_fine}
\end{figure}

We compare our model predictions with the SDSS measurements presented by \cite{Zehavi11}.
In the top panel of Figure \ref{wp_fine}, we show the results from the no-scatter model by the dashed lines.
The adopted sphere size to measure local DM densities around subhalos is same as the best size to reproduce the observed red and blue galaxy correlation functions, i.e., $R_{\rm DM}=200~\hkpc$ (see Section \ref{sec:wp}).
Note that the three inner most bins are not reliable since our SHAM implementation does not recover the observed projected correlation functions for the full sample of the faintest magnitude sample at that scales (see Figure \ref{wp_hod_lum}).
The figure shows that the no-scatter local density model gives nice agreements between the SDSS measurements and the model results for the reddest and bluest samples although the agreements for the redder and bluer ones are less impressive.
It would be because that the one-to-one relation for subhalo local density and galaxy color is too strong and should be more moderate.

We expect that introducing the scatter in the local DM density-color relation improves the agreements between the SDSS observation and our local density model.
As in the previous section, we use the perturbed local DM density around subhalos rather than the originally measured density and search the best combination of $\sigma_{\rm p}$ and $R_{\rm DM}$ to reproduce the observed projected correlation functions.
To evaluate matching of our model with scatter to the SDSS correlation functions, we calculate the $\chi^2$ value for each combination as
\begin{equation}
  \chi^2=(\chi^2_{\rm reddest}+\chi^2_{\rm redder}+\chi^2_{\rm bluer}+\chi^2_{\rm bluest})/{\rm d.o.f.}.
\end{equation}
To calculate the chi-square value of each fine color bin, we use Equation (\ref{eq:chi2}) but we do not include the three inner most bins.

The bottom panel of Figure \ref{wp_fine} shows the chi-square value $\chi^2$ as a function of $\sigma_{\rm p}$ and $R_{\rm DM}$.
We find that the combination of 
\begin{equation}
  \sigma_{\rm p}=0.3~~{\rm and}~~R_{\rm DM}=750~\hkpc
\end{equation}
gives the least $\chi^2$ value of $3.3$ among our twenty-four combinations.
The top panel shows the result for each color bin from the best combination by the solid lines.
It is clearly seen that the scatter model matches the observed correlation functions better than the no-scatter model.
In particular, the improvement in the redder sample is impressive.
It should be noted that the bottom panel represents other combinations give very similar $\chi^2$ values to that of the best combination.
For instance, we have the second least chi-square value of $\chi^2=3.4$ if we set $\sigma_{\rm p}=0.2$ and $R_{\rm DM}=500~\hkpc$.
It means that the projected correlation functions in finer color bins alone do not make tight constraint on the perturbation parameter $\sigma_{\rm p}$ and the local density measure $R_{\rm DM}$.

\sm{\subsubsection{Red fraction profiles in galaxy groups and clusters}}
In the previous section, we have shown that introducing finite scatter in the local DM density-color relation 
is clearly needed to explain the observed correlation functions in the finer color bins.
However, the parameter degeneracies in the scatter model calls for the need of additional observations to better constrain $R_{\rm DM}$ and $\sigma_{\rm p}$.
It might be better to use the lensing measurements for the finer color bins  together with the correlation functions as we did in Section \ref{del_col}.
However, unfortunately, we do not have such observational results yet.

As an alternative, we consider the red galaxy fraction profile in galaxy group- and cluster-sized halos \citep[see e.g.,][]{Hansen09,Wetzel12}.
Here the red and blue galaxies are divided by Equation (\ref{divider}) and include the reddest and redder, and the bluest and bluer subsamples, respectively \citep[see][]{Zehavi11}.
For this purpose, we use only the magnitude bin sample of $-20<M_r<-19$, locating within halos with masses greater than $10^{13}\hMsun$.
We stack the red and blue galaxy distributions in all corresponding halos to obtain the average red fraction as a function of the spatial distance from the halo center in units of the pseudo-virial radius $R_{200}$\footnote{The internal density within the spherical bound region with the radius $R_{200}$ is $200$ times higher than the critical density of the Universe.}.
The halo center is defined to be the density maximum of its smooth component.

\begin{figure}
  \includegraphics[width=8.5cm]{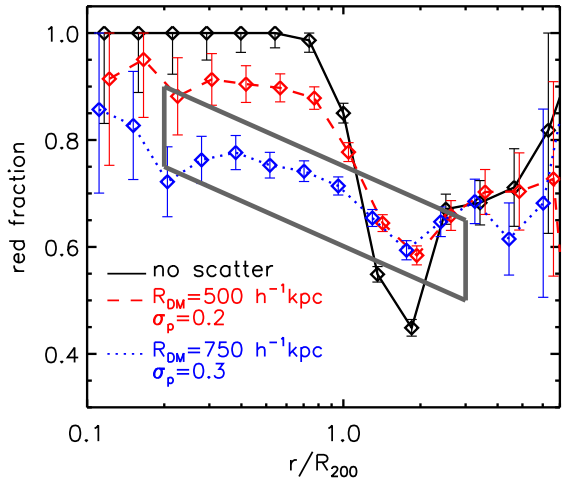}
  \caption{The average red fraction profiles of the galaxies with $-20<M_r<-19$ as a function of the spatial distance from the halo center in units of the pseudo-virial radius $R_{200}$. The solid line shows the result from the no-scatter local density model with $R_{\rm DM}=200~\hkpc$. Other lines, the dashed and dotted ones, represent the results from the scatter models with $(R_{\rm DM}/[\hkpc],~\sigma_{\rm p})=(500,~0.2)$ and $(750,~0.3)$, respectively. The error bars are computed assuming Poisson statistics. Note that the two scatter models give very similar chi-square values for the correlation functions in the finer color bins. The grey box shows very crudely the results by \citet{Hansen09}.}
  \label{red_frac}
\end{figure}

We show the resulting average profiles in Figure \ref{red_frac}.
The figure shows that the no-scatter model predicts the unnatural fraction values of unity at $r<0.5R_{200}$.
This is just because that the no-scatter model simply paints subhalos in high density regions red.
On the other hand, the two scatter models give more moderate values at $r<R_{200}$ since introducing scatter softens the local density-color relation.
Moreover the results from the scatter models with $(R_{\rm DM}/[\hkpc],~\sigma_{\rm p})=(500,~0.2)$ and $(750,~0.3)$ show distinguishable differences among them at $r<R_{200}$.
The former model predicts values of $\simeq0.9$ and a flat feature while the latter one's profile has lower values and is decreasing for increasing $r$.
It should be noted that the chi-square values from the two scatter models for the projected correlation functions in the finer color bins are very similar: the model with $R_{\rm DM}=500~\hkpc$ and $\sigma_{\rm p}=0.2$ gives $\chi^2=3.4$ and the other does $\chi^2=3.3$.
This relatively strong dependence of the red fraction profile on the parameters $R_{\rm DM}$ and $\sigma_{\rm p}$ is due to the fact that the red fraction profile is similar to the relative bias of red and blue galaxies in the one-halo term regime.
The relative bias enhances the small dependences of galaxy correlation functions in color bins on the parameters $R_{\rm DM}$ and $\sigma_{\rm p}$.
Hence, as well as correlation functions and lensing profiles, the red fraction profiles in galaxy groups and clusters would be an important clue to study the subhalo property-galaxy color relations.

The grey box in Figure \ref{red_frac} shows approximately the red fraction profile measured by \cite{Hansen09}.
\cite{Hansen09} classified clusters by richness and measured the red fractions for galaxies with $^{0.25}M_i-5\log_{10}h<-19$ in bins of richness, where $^{0.25}M_i$ is the $i$-band luminosity K-corrected to $z=0.25$.
They showed that red fraction profiles in very rich clusters have higher values than those in poor ones.
They also found that the red fraction is correlated with cluster richness for poor clusters with richness within $R_{200}$ of $N_{200}<10$, but is independent of richness for richer systems.
Hence the height of the grey box in the figure stands for the richness dependence of the red fraction.
The figure shows that the red fraction from the scatter model with $R_{\rm DM}=750\hkpc$ and $\sigma_{\rm p}=0.3$ is within the observational results.
Since the galaxy sample used by \cite{Hansen09} is not same as \cite{Zehavi11}, and we do not follow cluster classification in \cite{Hansen09}, we here cannot discuss consistency between our model predictions and the observed fraction quantitatively as we did for correlation functions.
Clearly it is important to study galaxy evolution and its correlation with DM structure evolution with several different observational measurements.

\section{Summary and Discussion}
We have extended the subhalo abundance matching (SHAM)  
to develop a new scheme that assigns galaxy color as well as luminosity to subhalos.
We consider the subhalo age and the local density 
as a secondary subhalo property which is expected to be correlated 
with galaxy color.
Technically, we divide a magnitude-binned subhalo catalog 
obtained by SHAM into two samples by the secondary property.
The two samples are then meant to represent red galaxies and blue ones.
In a similar fashion to SHAM, the abundance ratio of the red and blue galaxies 
is matched to the observed ratio.
We have studied the spatial clustering of the red and blue galaxies
based on the subhalo age model and the local density model.
Overall, the two models reproduce the observed color-dependent galaxy clustering properties \citep{Zehavi11} reasonably well. It is encouraging that SHAM can be
extended successfully in this way by using a secondary subhalo property
that can be easily measured.

We have examined a few other subhalo properties as an alternative proxy for galaxy color,
for instance, the density concentration and the spin parameter.
We have also tried with slightly different definitions for the concentration, spin, 
age and local DM density.
None of them yields better agreement with the observed correlation functions.
We thus conclude that our original age model and the local density model are the best 
among those we have examined.
It is worth discussing further the success of our local density model. 
One can naively expect that galaxies in higher density environment 
formed earlier and that they are not forming stars actively any more. 
Thus the local matter density could be related to the galaxy color.
The local DM density around a subhalo within $100~\hkpc$-order sphere \sm{would}
closely trace the subhalo mass\sm{, especially for central galaxies}.
Then the subhalo mass itself could be used as a proxy for the galaxy color rather than the local density.
We can test the idea by using the subhalo masses determined by the {\it SubFind} algorithm 
at the observation time.
In this model, more (less) massive subhalos correspond simply to redder (bluer) galaxies 
similarly to the local density model.
We have run the same calculations in the previous sections. 
\sm{We have found that the HODs of red/blue central galaxies from this subhalo mass model are similar to those from the local density model.
On the other hand, in this subhalo mass model, the HODs of red satellite galaxies have lower amplitude than those of blue satellite galaxies at overall halo mass range in contrast to the local density model.}
Because \sm{satellite} subhalos in a massive distinct halo typically have small masses, 
owing to tidal mass stripping, they tend to be assigned a blue galaxy in this model.
\sm{Then} the subhalo mass model does not reproduce at all the observed color-dependent clustering.
Indeed, the selected blue galaxies show strong clustering at all length scales.
We argue that some environmental effect, in terms of local density, is important for 
color assignment for the satellite galaxies.

It is crucial to test our model predictions against not only spatial clustering
but also other measurement(s). We propose to use the color dependence of lensing profile, 
motivated by the observed morphology dependence \citep{Mandelbaum06}.
To this end, we regard the early (late) types as the red (blue) galaxies
and calculate the lensing profiles for the two populations.
The predictions from our two models 
agree with the observed red galaxy lensing profile reasonably well.
In particular the local density model reproduces the observed
lensing profile for the blue (late type) galaxies.
However, while the local density model shows a large color (morphology) split 
at $100~\hkpc\la r_p\la1~\hMpc$,
the age model does not account for the low amplitude 
of the lensing profiles of the late (blue) galaxies with $-21<M_r<-20$ 
and $-22<M_r<-21$ at $100~\hkpc\la r_p \la1~\hMpc$ at all.

 The fact that the age model does not reproduce 
the observed color (morphology) dependence of lensing profiles
might indicate that galaxy color and subhalo age are not tightly correlated with each other.
There is an observational hint for this notion.
\cite{Tinker11} studied the fraction of central galaxies that are not forming stars actively
(``quenched centrals'') as a function of 
the local galaxy density using a SDSS group catalog.
They show that the fraction of passive galaxies does not agree with the fraction of old halos,
in particular for relatively small galaxies with $\log[M_*/(h^{-2}M_\odot)]<10$.
Their findings are qualitatively consistent with our result.
There is not a tight relation between the galaxy color and the age of the host (sub)halo,
or one may need to devise a rather complex conversion from one to the other.
We can think of using the subhalo assembly history of red and blue galaxies directly obtained 
from $N$-body simulations. Further studies are clearly needed.

\sm{Despite of the reasonable success of our local density model, the clustering amplitudes of 
the blue galaxies in the local density model 
is systematically smaller than the observation 
at $r_p\la20~\hkpc$, where the central-satellite signal dominates.
This is because, in the local density model, 
central blue galaxies tend to be hosted by relatively small distinct halos (see Figure \ref{hod_col}).
The number of pairs of a blue central and a blue satellite 
is then smaller than in the observation.
It is expected that the small-scale clustering can be better matched to the observation 
by introducing scatter in the local DM density-color relation.
In the real universe, there are substantial scatters in the relation.
We have modeled the scatters by perturbing the measured local DM density 
with a simple probability distribution, e.g., a Gaussian with a finite variance.
Such an `improved' model would then assign a blue central galaxy more likely 
to a massive distinct halo.
We have shown that introducing scatters in the local density - color
relation, in the above {\it ad hoc} manner, indeed improves the agreement in 
the small-scale clustering of the blue galaxies (see Figure \ref{wp_hod_sca}). 
Interestingly, the best sphere size for measuring local DM densities
becomes  larger than in our original local density model. 
Essentially, introducing scatter is similar to averaging over the red and blue 
subhalo samples.
Then the ``color-split'' in the two-point correlation functions becomes smaller.
Hence the best sphere size needs to be  larger in order for a model
with scatters to reproduce the observed clustering 
(see Figure \ref{wp_hod_loc}).
However it should be noted that our treatment for introducing scatter is very crude.
It is  necessary to construct methods to include scatter in physically motivated ways.}

\sm{We  have also studied the correlation functions in the finer color bins.
It is natural to expect scatters in the local DM density-color relation to play more important roles in finer color binned samples.
We have shown that the local density model without scatter fails to explain the observed clustering in the finer color bins and that introducing scatter is needed to reproduce the observations (see Figure \ref{wp_fine}).
However we also have shown that using the correlation functions in the finer color bins alone is not enough  to constrain our model parameters.
We have explored dependences of the red galaxy fraction profiles in galaxy groups and clusters on the parameters.
We have found that the red fraction profile depends on the parameters relatively strongly (see Figure \ref{red_frac}).
Therefore, as well as spatial clustering and galaxy-galaxy lensing, the red galaxy fraction can be regarded as a promising observable clue to study the subhalo property-galaxy color relations.
Recently \cite{Hearin12} argued that the SHAM prescription has a difficulty to reproduce the observed galaxy group multiplicity function and spatial clustering simultaneously.
Together with our findings in this paper, the galaxy groups and clusters statistics should be concerned to study the galaxy-DM (sub)halo connections.
These statistics may also prove important for extending our methods to higher redshifts where lensing-related measurements become more challenging.}

The abundance matching is a powerful technique to populate a cosmological $N$-body simulation
with galaxies. It is encouraging that our simple models reproduce the observed color-dependent
clustering well. There are a variety of choices of the secondary subhalo property as a proxy 
of galaxy color, and thus it is important to test models using other observations independent of
spatial clustering, as we have done in the present paper.

\section*{ACKNOWLEDGMENTS}
We thank Steven Bamford, Shirley Ho, Eiichiro Komatsu, Alexie Leauthaud, Surhud More, Kentaro Nagamine and Naoshi Sugiyama for useful discussions and suggestions.
We appreciate Rachel Mandelbaum, Erin Sheldon, Idit Zehavi and Zheng Zheng for sending us their data, and Takahiro Nishimichi for providing us with the second-order Lagrangian perturbation theory code.
We are grateful to the anonymous referee for helpful comments.
S.M. is supported by the JSPS Research Fellowship.
Y.T.L. acknowledges supports from the World Premier International Research Center Initiative, MEXT, Japan \fig{and the National Science Council grant NSC 102-2112-M-001-001-MY3.}
N.Y. acknowledges financial support by the Grant-in-Aid for Young Scientists　(S 20674003) by JSPS. 
The work is supported in part by Kobayashi-Maskawa Institute and Global COE at Nagoya University, by WPI Initiative by MEXT, and by Grant-in-Aid for Scientific Research on Priority Areas No. 467.

\bsp

\label{lastpage}

\end{document}